\documentclass[12pt,peerreview,draftcls]{IEEEtran}
\usepackage{mathrsfs}
\usepackage{bbding}
\usepackage{graphicx}
\usepackage{float}
\usepackage{booktabs}
\usepackage{hyperref}
\usepackage{cite}

\usepackage{enumitem}
\usepackage{color}
\usepackage{ulem}
\usepackage{psfrag}
\usepackage{subfigure}
\usepackage{amssymb}
\usepackage{amsthm}
\usepackage{setspace}
\usepackage{epsfig}
\usepackage{pifont}
\usepackage{amsmath}
\usepackage{array}
\newcommand{\PreserveBackslash}[1]{\let\temp=\\#1\let\\=\temp}
\newcolumntype{C}[1]{>{\PreserveBackslash\centering}p{#1}}
\newcolumntype{R}[1]{>{\PreserveBackslash\raggedleft}p{#1}}
\newcolumntype{L}[1]{>{\PreserveBackslash\raggedright}p{#1}}
\usepackage{multicol}
\usepackage{multirow}
\usepackage{diagbox}
\usepackage{pifont}
\usepackage{indentfirst}
\usepackage{amsfonts}
\usepackage{algorithm}
\usepackage{algorithmicx}
\usepackage{algpseudocode}
\floatname{algorithm}{Procedure}
\usepackage{fancyhdr}
\usepackage{amscd}
\usepackage[cmyk]{xcolor}
\usepackage{bm}
\usepackage{fancyhdr}
\setlist{itemsep=0pt,parsep=0pt}

\renewcommand{\arraystretch}{1.2}
\newtheorem{proposition}{Proposition}
\newtheorem{definition}{Definition}

\newtheorem{remark}{Remark}
\newcommand{\tabincell}[2]{\begin{tabular}{@{}#1@{}}#2\end{tabular}}
\renewcommand{\arraystretch}{1.1} % 调整表格内单元格上下行距
\allowdisplaybreaks[4]
\IEEEoverridecommandlockouts

\renewcommand{\arraystretch}{0.9}	
\usepackage{caption}
\begin{document}
	\title{Recursive GNNs for Learning Precoding Policies with Size-Generalizability}
	
	\author{
		% \thanks{This work is supported by National Natural Science Foundation of China (NSFC) under Grant XXX}
		\IEEEauthorblockN{Jia Guo and Chenyang Yang}
\thanks{A part of this work has been published in IEEE VTC Fall 2023\cite{VTCFGJ2023}.}.
		% \thanks{If this paper is accepted, we will publicize our codes on GitHub.}
		%	
		
		\IEEEauthorblockA{Beihang University, Beijing, China, Email: \{guojia, cyyang\}@buaa.edu.cn}}
	\maketitle
	\setcounter{page}{1}
	\thispagestyle{empty}
	
	\begin{abstract}
		Graph neural networks (GNNs) have been shown promising in optimizing power allocation and link scheduling with good size generalizability and low training complexity. These merits are important for learning wireless policies under dynamic environments, which partially come from the matched permutation equivariance (PE) properties of the GNNs to the policies to be learned.
		Nonetheless, it has been noticed in literature that only satisfying the PE property of a precoding policy in multi-antenna systems cannot ensure a GNN  for learning precoding to be generalizable to the unseen number of users.
		Incorporating models with GNNs helps improve size generalizability, which however is only applicable to specific problems, settings, and algorithms.
		In this paper, we propose a framework of size generalizable GNNs for learning precoding policies that are purely data-driven and can learn wireless policies including but not limited to baseband and hybrid precoding in multi-user multi-antenna systems. To this end, we first find a special structure of each iteration of two numerical algorithms for optimizing precoding, from which we identify the key characteristics of a GNN that affect its size generalizability. Then, we design size-generalizable GNNs that are with these key characteristics and satisfy the PE properties of precoding policies in a recursive manner. Simulation results show that the proposed GNNs can be well-generalized to the number of users for learning baseband and hybrid precoding policies and require much fewer samples than existing counterparts to achieve the same performance.
		
		\begin{IEEEkeywords}
			Graph neural networks, size generalizability, precoding, permutation equivariance
		\end{IEEEkeywords}
	\end{abstract}
	
\vspace{-4mm}\section{Introduction}
	Learning precoding policies with deep neural networks (DNNs) enables real-time implementation  \cite{FNN_mUE_DigPre_WCL_2020_Kim}, joint optimization with channel acquisition \cite{GNN_mUE_RIS_JSAC_2021_Yu}, and robustness to channel estimation errors \cite{CNN_sUE_HyPre_TCOM2022_Liu}.
Most existing works designed fully-connected neural networks (FNNs) \cite{FNN_mUE_DigPre_TVT2021_Kong, FNN_mUE_DigPre_WCL_2020_Kim} and convolutional neural networks (CNNs) \cite{CNN_sUE_HyPre_TCOM2022_Liu, CNN_sUE_HyPre_TWC2020_Peken, DL_mBS_DigPre_GC2020_Lee, DL_Beamforming_ModelBased_TWC_2021}, %Model_DL_HybPrec_TCOM2023, DL_Beamforming_model_TCOM2020, DL_Precod_Opt_Struc_TCOM2023}
 for learning precoding policies, which can achieve satisfactory performance for a given problem scale (e.g., for a given number of users). However, precoding is often used in dynamic systems where the channel statistics and the problem scale change over time.
It has been observed that DNNs can be adaptive to non-stationary channels by resorting to  continuous learning \cite{L2ContinueOpt} or meta-learning \cite{Meta_Beamforming_TWC2021}. Yet both FNNs and CNNs lack the generalizability to their input sizes, even after using meta-learning \cite{ZBC_Meta_Learning_GC2023}.
	
Recently, graph neural networks (GNNs) have been designed to learn wireless policies such as power control \cite{Eisen2020, shen2019graph}, power allocation \cite{GJ_TWC_GNN}, link scheduling \cite{lee2020wireless} and precoding \cite{GNN_mUE_RIS_JSAC_2021_Yu,GNN_Beamforming_TCOM2023,ZBC_WCNC,ZBC_VTC2023,LSJ_MultiDim_GNN_2022,MCast_UCast_Beamform_TCOM2023,GAT_precoding_2023}. Their generalizability to unseen graph sizes \cite{Eisen2020, shen2019graph, GJ_TWC_GNN,lee2020wireless}, scalability to large scale systems \cite{Eisen2020, shen2019graph,lee2020wireless}, and higher sample efficiency \cite{Eisen2020, shen2019graph,GJ_TWC_GNN,ZBC_WCNC,ZBC_VTC2023} have been demonstrated by extensive simulations.
For example, the GNNs for learning the power control/allocation and link scheduling policies \cite{Eisen2020,shen2019graph, GJ_TWC_GNN,lee2020wireless} are adaptable to different input sizes without the need of re-training or fine-tuning. It has been noticed that these advantages of GNNs stem from designing their parameter sharing schemes by harnessing the permutation equivariance (PE) properties \cite{Eisen2020,shen2019graph,GJ_TWC_GNN}, which widely exist in wireless policies  \cite{LSJ_MultiDim_GNN_2022}. Nonetheless, existing GNNs for learning precoding policies cannot be well-generalized to the number of users, even if the GNNs satisfy the PE properties of the policies. In particular, the GNNs for learning precoding policies in multi-user multi-antenna systems perform well only when the number of users in the test set is close to or smaller than that in the training set \cite{MCast_UCast_Beamform_TCOM2023,GAT_precoding_2023}, or when the signal-to-noise (SNR) is low \cite{GNN_mUE_RIS_JSAC_2021_Yu}.
How to design GNNs for learning precoding policies that are generalizable to large numbers of users in various settings remains open.
	
\vspace{-2mm}\subsection{Related Works}\vspace{-1mm}
\subsubsection{Learning to Optimize Precoding Policies}
	Deep learning has been introduced to optimize precoding in various multi-antenna systems for reducing inference time \cite{FNN_mUE_DigPre_WCL_2020_Kim, ZBC_WCNC, Meta_Beamforming_TWC2021,LSJ_MultiDim_GNN_2022}, jointly optimizing with channel acquisition \cite{FNN_mUE_DigPre_TVT2021_Kong, GNN_mUE_RIS_JSAC_2021_Yu, DL_ChlEst_Precod_TWC_2021}, or improving the robustness to channel estimation errors \cite{CNN_sUE_HyPre_TCOM2022_Liu,MCast_UCast_Beamform_TCOM2023}. For example, in \cite{FNN_mUE_DigPre_TVT2021_Kong, FNN_mUE_DigPre_WCL_2020_Kim}, FNNs were used to learn baseband precoding policies in multi-user multi-antenna systems, where the FNN was trained in unsupervised manner  in \cite{FNN_mUE_DigPre_WCL_2020_Kim} to avoid generating labeled samples. In \cite{CNN_sUE_HyPre_TCOM2022_Liu, CNN_sUE_HyPre_TWC2020_Peken}, CNNs were designed to learn single-user hybrid precoding policies for millimeter-wave (mmWave) communication systems to exploit the spatial correlation of channel matrices. In \cite{DL_mBS_DigPre_GC2020_Lee}, a CNN was designed to learn a coordinated beamforming policy in a multi-cell multi-antenna system.
	
Both CNNs and FNNs are not embedded with any PE property, while all precoding policies in different system settings are with permutation properties \cite{LSJ_MultiDim_GNN_2022}. Due to the large hypothesis space, these DNNs require high training complexity \cite{ZBC_WCNC}. For example, the FNN applied in \cite{FNN_mUE_DigPre_WCL_2020_Kim} needs to be trained with 10,000 samples even for a small-scale problem with six antennas at the base station (BS) and six users. Due to not satisfying permutation property, these DNNs cannot be generalized to problem scales.
	
	To show the gain from harnessing the PE property of precoding, a GNN was designed and compared with CNNs in \cite{ZBC_WCNC} to learn baseband precoding policy in multi-user multi-antenna systems. In \cite{GNN_mUE_RIS_JSAC_2021_Yu,ZBC_VTC2023}, GNNs were designed to optimize beamforming and reflective coefficients in intelligent reflecting surface-aided multi-antenna systems. While the PE properties of the precoding policies have been harnessed by the GNNs designed in \cite{ZBC_WCNC, GNN_mUE_RIS_JSAC_2021_Yu,ZBC_VTC2023}, the training complexity of the GNNs are still high, albeit much lower than that of the CNNs, and the GNNs cannot be generalized to  the number of users unless the SNR is very low \cite{GNN_mUE_RIS_JSAC_2021_Yu,ZBC_WCNC}.
	
	% All these works learn the spectral efficiency (SE)-maximal precoding policies, except that the precoding for maximizing the minimal data rate was also considered in \cite{GNN_mUE_RIS_JSAC_2021_Yu}.
	
Maybe realized that the DNNs for directly learning the mapping from channel matrix to precoding matrix is hard to train, most research efforts for optimizing precoding turn to design model-driven DNNs \cite{DeepUnfold_WMMSE_TWC_2021,GNN_Beamforming_TCOM2023,Model_DL_HybPrec_TCOM2023, DL_Beamforming_model_TCOM2020, ModelGNN_GJ_2022,DL_Precod_Opt_Struc_TCOM2023}. The models incorporated into DNNs can be iterative algorithms \cite{Model_DL_HybPrec_TCOM2023,DeepUnfold_WMMSE_TWC_2021,ModelGNN_GJ_2022}, or the structures of optimal baseband precoding matrices of several optimization problems \cite{GNN_Beamforming_TCOM2023,DL_Beamforming_model_TCOM2020,DL_Precod_Opt_Struc_TCOM2023}. Although the training complexity can be reduced, the applications of the model-driven DNNs are restricted by the introduced models. For example, the beamforming neural network proposed in \cite{DL_Beamforming_model_TCOM2020} and the deep unfolding network proposed in \cite{DeepUnfold_WMMSE_TWC_2021}
	that respectively leverage the optimal baseband precoding matrix structure and the weighted-minimum-mean-squared-error (WMMSE) algorithm \cite{WMMSE2011Shi}, and hence
	cannot be used to learn hybrid precoding policies.
	
	\subsubsection{Improving Size-generalizability of GNNs}
In order for enabling size-generalizability, the architectures and training algorithms of GNNs have been  designed in the field of machine learning. The architecture of a GNN can be characterized by its update equation, which is used to update the hidden representations of each vertex or edge of a graph in each layer. Specifically, the information of neighbored vertices and edges is firstly processed by a {processor} and then aggregated by a pooling function. Then, the aggregated information is combined with the previous representation of the vertex or edge itself with a {combiner}.
In \cite{corso2020principal}, an aggregation function of GNN was designed to aggregate more information from neighbored vertices. In \cite{knyazev2019understanding,velivckovic2017graph}, attention mechanism was introduced into the GNN architecture for only aggregating important information from neighboring vertices. The designed GNNs in \cite{corso2020principal,knyazev2019understanding,velivckovic2017graph} were shown to be  generalizable to problem scales via empirical evaluations.
	In \cite{fu2021generalize}, a GNN was pre-trained and fine-tuned respectively in small and large problem scales to improve its performance under different problem scales. In \cite{tang2020towards}, a GNN was designed, whose number of hidden layers changes with the problem scale and can be controlled by learning a terminate condition.
In \cite{battaglia2016interaction,li2022ood,yehudai2021local}, the loss functions for training GNNs were designed, where semi-supervised learning was adopted in \cite{yehudai2021local} to learn the common characteristics in both the small- and large-scale problems.

	In the field of wireless communications, most previous works  \cite{Eisen2020, shen2019graph, GNN_mUE_RIS_JSAC_2021_Yu, GNN_Beamforming_TCOM2023, GJ_TWC_GNN, lee2020wireless, GAT_precoding_2023,MCast_UCast_Beamform_TCOM2023} adopted the GNNs with well-established architectures and evaluated the size generalizability of these GNNs for learning wireless policies via simulations. In \cite{WJJ_GNN_Gen},  the size generalizability mechanism of GNNs  for learning the policies with one-dimensional (1D)-PE property was analyzed, and a GNN with pre-trained activation function in the output layer was proposed to enable the generalizability to the number of users for learning power and bandwidth allocation policies.
	
	The size generalizability of GNNs depends on the tasks to be learned. For example, the GNNs in \cite{Eisen2020, shen2019graph,GJ_TWC_GNN,lee2020wireless} that can be well-generalized to problem scales for learning power control/allocation or link scheduling policies are not size-generalizable for learning precoding policies. As to be analyzed later, a GNN for learning precoding policies cannot be well-generalized to the number of users unless a processor in the GNN is judiciously designed.
	%unless a processing and a combination functions in the GNN are judiciously designed such that the learned functions are identical for each problem scale.
	
%	for learning a policy such as precoding, a GNN can be generalized to problem scales only if a processing and a combination functions in the GNN are judiciously designed such that they are independent of problem scales.
%	Despite the methods proposed in the above-mentioned works improve the size generalizability in their considered tasks, the designs of GNN structures and learning algorithms are not guaranteed to improve size generalizability when learning wireless policies.
	% A, \cite{WJJ_GNN_Gen} is the only work that theoretically analyzes the size generalizability of GNN for learning wireless policies,  However, the size generalizability for learning precoding was not considered in \cite{WJJ_GNN_Gen}.
	% Besides, an underlying assumption in the works of designing learning algorithms \cite{fu2021generalize,battaglia2016interaction,li2022ood,yehudai2021local} is that a size-generalizable GNN structure has already been well-designed such that only the optimal trainable parameters need to be found, which however is not the case for learning precoding policies.

\vspace{-4mm}\subsection{Motivations and Contributions}\vspace{-1mm}
	In this paper, we propose a framework of designing size-generalizable GNNs for learning precoding policies.
%The update equation is equivariant to the permutations of vertices because the processor and combiner are respectively identical for all the vertices.
We focus on the generalizability to the number of users, which is harder than improving the generalizability to the number of antennas \cite{LSJ_MultiDim_GNN_2022}.

Inspired by the observation in \cite{WJJ_GNN_Gen} (i.e., the size scaling law of a size-generalizable GNN should be aligned with the size scaling law of a policy), we design the update equations of GNNs by analyzing the numerical algorithms for optimizing precoding. Since most optimal precoding matrices do not have closed-form expressions, the size scaling law of a precoding policy is hard to find if not impossible.
Noticing that an approximate precoding policy can be obtained from a numerical algorithm whereas the algorithm with multiple iterations makes the approximate policy not with a simple structure, we examine how the input-output relation of each iteration of two numerical algorithms for optimizing precoding scales with sizes. We find that the input-output relation can be expressed as a function with special structure that satisfies 1D-PE property, and the iterative equation obtained by re-expressing the iteration of an algorithm can be regarded as the update equation of a well-trained GNN (called WGNN) with the 1D-PE property.
The processor and combiner of the update equation of the WGNN are  size-invariant, i.e., independent of problem scales. To design the GNN architecture with size-generalizability, we first identify three essential characteristics that lead to the  size-invariance of the WGNN.
 %by comparing the update equations of existing non-size-generalizability GNNs with the WGNN.
Then, we design the GNNs with these characteristics, whose update equations are with the same structure as that of the WGNN.

	Since both the features and the actions are defined on edges of the graphs for precoding, edge-GNNs are used to learn the precoding policy, where the hidden representation of every edge is updated in each layer of the GNN.
	The main contributions are summarized as follows.
	\begin{itemize}
		\item We find that for optimizing spectral efficiency (SE)-maximal baseband precoding and hybrid precoding, the input-output relation of each iteration of two numerical algorithms can be expressed as a function satisfying 1D-PE property with a special structure, referred to as a \emph{1D-PE function}.
%, and the iteration equations of each algorithm can be regarded as the update equation of a well-trained size-generalizable GNN.
From the special structure, we discover essential characteristics that enable a GNN to learn precoding policies with size generalizability. Specifically, to update the hidden representations of the edges connected to a user, a non-linear processor should be used to capture the correlation between the representations of these edges and the representations of the edges connecting to other users, and should not process the representation of each edge independently.
%		\item We find that when updating the hidden representations of each edge in each layer of a GNN for precoding that is permutation
%equivariant to users, the processor should be with the following characteristics for enabling the generalizability to the number of users. Specifically, to update the hidden representations of the edges connected to a user, a non-linear processor should be used to capture the correlation between the representations of these edges and the representations of the edges connecting to other users, and should not process the representation of each edge independently.
		\item We propose a GNN architecture with these characteristics, which satisfies high-dimensional permutation property of a precoding policy in a recursive manner and hence is called recursive GNN (RGNN). The update equations of the RGNNs are with the special structure of the \emph{1D-PE function}, which resemble the re-expressed iterative equations of numerical algorithms for precoding. Different from the model-driven DNNs that simplify the mapping to
be learned from data by incorporating with particular models, the RGNNs restrict the hypothesis space by leveraging the common structure of numerical algorithms hence can be applied to learn policies including but not limited to baseband and hybrid precoding.
		\item We validate the generalizability of the RGNNs via simulations. The results show that the RGNNs can be well-generalized to the number of users for learning SE- and energy efficiency (EE)-maximal baseband precoding and SE-maximal hybrid precoding policies, and require lower training complexity to achieve an expected performance.
	\end{itemize}

While a part of this work has been published in \cite{VTCFGJ2023}, the contents in this journal version are substantially extended, including finding the common structure of each iteration of numerical algorithms for both baseband and hybrid precoding problems, analyzing sample and inference complexity, and simulations results for
EE-maximal precoding.
	
	\emph{Notations}: $(\cdot)^{\sf T}$ and $(\cdot)^{\sf H}$  respectively denote  transpose and conjugate transpose. $\|\cdot\|$ and $\|\cdot\|_F$ denote two-norm and Frobenius norm. $|\cdot|$ denotes the magnitude of a complex value.
	% ${\bf X}=[x_{ij}]^{K\times N}$ denotes a $K\times N$ matrix, and its element in the $i$-th row and $j$-th column is $x_{ij}$,
	$({\bf X})_{i,j}$ denotes the element in the $i$-th row and $j$-th column of matrix ${\bf X}$.
	
	The rest of the paper is organized as follows. In section \ref{sec: system model} and section \ref{sec:vgnn}, we respectively introduce the precoding policies and the existing GNN architectures for optimizing precoding. In section \ref{sec: rgnn design}, we identify the key characteristics for enabling generalizability. In section \ref{sec:rgnn-design}, we design RGNNs by incorporating these characteristics. In section \ref{sec:simulation-results}, we provide simulation results to compare the generalizability of the GNNs to the number of users, as well as their training complexity. In section \ref{sec:conclusions}, we provide the conclusion remarks.
	
\vspace{-3mm}\section{Precoding Policies and Their Properties}\label{sec: system model}
	In this section, we introduce the optimization problems for precoding and the resulting policies.

 	Consider a downlink multi-user multi-antenna system, where a BS equipped with $N$ antennas transmits to $K$ users each with a single antenna. For easy exposition, we take the SE-maximal bandband and hybrid precoding problems as two use cases, which are widely investigated challenging problems. Nonetheless, the permutation properties as well as the designed GNNs are also applicable to the precoding problems with other objective functions and constraints.	
%\vspace{-3mm}\subsection{Baseband Precoding}

 	The baseband precoding at the BS can be optimized by maximizing the SE as follows,
 	\begin{subequations}
 		\begin{align}
 		{\bf P1}:
 		\max_{{\mathbf{W}}_{BB}}~~ & \sum_{k=1}^K \log_2 \left(1+\frac{|{\mathbf{h}}^{\sf H}_k {\mathbf{w}}_{BB_k}|^2}{\sum_{i=1,i\neq k}^K  |{\mathbf{h}}^{\sf H}_k {\mathbf{w}}_{BB_i}|^2+\sigma_0^2} \right) \label{eq:bb-object} \\
 		{\mathrm{s.t.}} ~~
 		&  \|{\mathbf{W}}_{BB} \|^2_F = P_{\max},\label{eq:bb-constraint}
 		\end{align}
 	\end{subequations}
 	where ${\bf W}_{BB}=[{\bf w}_{BB_1},\cdots,{\bf w}_{BB_K}]\in {\mathbb C}^{N\times K}$ is the baseband precoding matrix, ${\bf w}_{BB_k}=[w_{BB_{1k}},\cdots,w_{BB_{Nk}}]^{\sf T}$ is the baseband precoding vector for the $k$-th user, ${\bf h}_k=[h_{1k},\cdots,h_{Nk}]^{\sf T}$ is the channel vector from the BS to the $k$-th user, $P_{\max}$ is the maximal transmit power at the BS, and $\sigma_0^2$ is the noise power. \eqref{eq:bb-constraint} is the power constraint.
 	
 	The baseband precoding policy is the mapping from the environmental parameters to the optimal precoding matrix ${\bf W}_{BB}^{\star}$. For notational simplicity, we only consider the channel matrix ${\bf H}\triangleq[{\bf h}_1,\cdots,{\bf h}_K]$ as the environmental parameter. Then, the baseband precoding policy is ${\mathbf{W}}_{BB}^{\star}=F_b({\mathbf{H}})$, where ${\mathbf{W}}_{BB}^{\star}$ is the optimal precoding matrix, and $F_b(\cdot)$ is a  multivariate function.
 	The policy satisfies the following two-dimensional (2D)-PE property, ${\bm \Pi}_{\mathsf{AN}}^{\mathsf{T}}{\mathbf{W}}_{BB}^{\star}{\bm \Pi}_{\mathsf{UE}}=F_b({\bm \Pi}_{\mathsf{AN}}^{\mathsf{T}}{\mathbf{H}}{\bm \Pi}_{\mathsf{UE}})$ \cite{ZBC_WCNC},
 	where ${\bm \Pi}_{\mathsf{AN}}$ and ${\bm \Pi}_{\mathsf{UE}}$ are arbitrary permutation matrices that respectively change the orders of antennas and users.
 	
%\vspace{-3mm}\subsection{Hybrid Precoding}
	To reduce the hardware complexity of massive multi-input-multi-output (MIMO) systems, the BS can also be equipped with $N_s$ RF chains and transmit with hybrid precoding.
	The baseband and analog precoding can be jointly optimized to maximize the SE as follows,
	\begin{subequations}
		\begin{align}
			{\bf P2}:
			\max_{{\mathbf{W}}_{RF},{\mathbf{W}}_{BB}} & \sum_{k=1}^K \log_2 \left(1+\frac{|{\mathbf{h}}^{\sf H}_k {\mathbf{W}}_{RF} {\mathbf{w}}_{BB_k}|^2}{\sum_{i=1,i\neq k}^K  |{\mathbf{h}}^{\sf H}_k {\mathbf{W}}_{RF} {\mathbf{w}}_{BB_i}|^2+\sigma_0^2} \right) \label{eq:obj} \\
			{\mathrm{s.t.}} ~~
			&  \|{\mathbf{W}}_{RF} {\mathbf{W}}_{BB} \|^2_F = P_{\max},\label{eq:constraint2}\\
			& |\left({\mathbf{W}}_{RF}\right)_{j,l}|=1,j=1,\cdots,N,l=1,\cdots,N_s\label{eq:constraint3},
		\end{align}
	\end{subequations}
	where ${\bf W}_{RF}\in{\mathbb C}^{N\times N_s}$ is the analog precoding matrix, \eqref{eq:constraint2} and \eqref{eq:constraint3} are respectively the power constraint and the constant modulus constraint of the analog precoder.
	
	Denote the hybrid precoding policy as $\{{\mathbf{W}}_{RF}^{\star},{\mathbf{W}}_{BB}^{\star}\}=F_h({\mathbf{H}})$, where ${\mathbf{W}}_{RF}^{\star}$ is the optimal analog precoding matrix, and $F_h(\cdot)$ is a  multivariate function.
	The policy satisfies the following three-set permutation property, $\{{\bm \Pi}_{\mathsf{RF}}^{\mathsf{T}}{\mathbf{W}}_{BB}^{\star}{\bm \Pi}_{\mathsf{UE}}, {\bm \Pi}_{\mathsf{AN}}^{\mathsf{T}}{\mathbf{W}}_{RF}^{\star}{\bm \Pi}_{\mathsf{RF}}\}=F_h({\bm \Pi}_{\mathsf{AN}}^{\mathsf{T}}{\mathbf{H}}{\bm \Pi}_{\mathsf{UE}})$ \cite{LSJ_MultiDim_GNN_2022},
	where ${\bm \Pi}_{\mathsf{RF}}$ is an arbitrary permutation matrix that changes the order of RF chains.

	\vspace{-1mm}
	\section{GNNs for Learning Precoding Policies}\label{sec:vgnn}
	\vspace{-1mm}	
	In this section, we recap two GNNs designed in \cite{ZBC_WCNC,LSJ_MultiDim_GNN_2022} and extend a graph attention network (GAT) designed in \cite{Edge-GAT} for learning the baseband and hybrid precoding policies. To facilitate the comparison with the numerical algorithms for solving precoding problems in the next section, we provide the update equations for the hidden representations of all the edges connected to each user of the GNN in \cite{ZBC_WCNC} and the extended GAT, and start by defining \emph{element-wise function} and \emph{non-element-wise function}.
	\begin{definition}
		For a multivariate function ${\bf y}=f({\bf x})$ that maps ${\bf x}\triangleq[x_1,\cdots, x_N]^{\sf T}$ to ${\bf y}\triangleq[y_1,\cdots, y_N]^{\sf T}$, if $y_n$ only depends on $x_n$, then $f(\cdot)$ is an element-wise function, i.e., ${\bf y}=f({\bf x})$ is the vector form of $N$ univariate functions. If $y_n$ depends not only  on $x_n$ but also on $x_i, i\neq n$, then $f(\cdot)$ is a  non-element-wise function.
	\end{definition}

\vspace{-7mm}	
\subsection{Baseband Precoding}\label{subsec:2d-vgnn}
\begin{figure}[!htb]
		\centering
		\includegraphics[width=.4\linewidth]{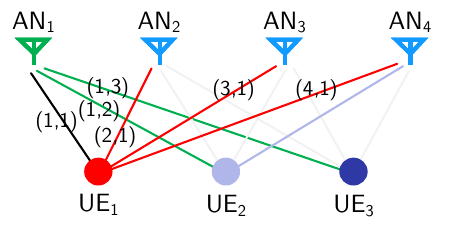}\vspace{-2mm}
		\caption{Example of the precoding graph for learning ${\bf W}_{BB}$, and the neighboring edges of edge $(1,1)$, $N=4$, $K=3$.}
		\label{fig:digital-precod}
	\end{figure}

\vspace{-1mm}	
	The precoding policy can be learned over a heterogeneous graph, which is composed of two types of vertices (i.e., user vertices and antenna vertices) and the edges connecting them \cite{ZBC_WCNC}. Denote the vertex corresponding to the $k$-th user as UE$_k$, and the vertex corresponding to the $n$-th antenna as AN$_n$. The vertices have no features or actions. The feature and action of the edge between AN$_n$ and UE$_k$, denoted as edge $(n,k)$, are respectively $h_{nk}$ and $w_{BB_{nk}}$.
%, which are assumed as real-valued scalar for notational simplicity.
As illustrated in Fig. \ref{fig:digital-precod}, edge $(1, 1)$ is connected to AN$_1$ with edge $(1, 2)$ and edge $(1,3)$, and is connected to UE$_1$ with edge $(2, 1)$, edge $(3,1)$ and edge $(4,1)$. We refer to edge $(1, 2)$ and edge $(1,3)$ as the \emph{neighboring edges} of edge $(1,1)$ by AN$_1$, and refer to edge $(2,1)$, edge $(3,1)$ and edge $(4,1)$ as the neighboring edges of edge $(1,1)$ by UE$_1$.
	
%	Since both the features and the actions are defined on edges, edge-GNNs can be used to learn the precoding policy, where the hidden representation of every edge is updated in each layer of the GNN.
	For notational simplicity and easy understanding, we assume that the hidden representation of each edge is a scalar, which is denoted as $d_{nk}^{(\ell)}$ for edge $(n,k)$ in the $\ell$-th layer. We also assume that $h_{nk}$ and $w_{BB_{nk}}$ are real-valued scalars. The assumptions will be removed later.
%The analysis in the sequel is still valid without these assumptions.
To obtain $d_{nk}^{(\ell)}$, the hidden representation of each neighboring edge of edge $(n,k)$ in the $(\ell-1)$-th layer is first passed through a processor for information extraction and then passed through a pooling function for information aggregation. The aggregated output is then combined with $d_{nk}^{(\ell-1)}$ to obtain $d_{nk}^{(\ell)}$. The inputs and outputs of such an edge-GNN with $L$ layers are respectively the features and actions of the edges, i.e., $d_{nk}^{(0)}=h_{nk}$ and $d_{nk}^{(L)}=w_{BB_{nk}}$.
	
	In the following, we provide the update equations of two edge-GNNs for learning the baseband precoding policy over the precoding graph, both adopting summation as the pooling function, but their processors and combiners are different.
	
	\subsubsection{2D-Vanilla-GNN}\label{sec:2d-vgnn}
	The edge-GNN designed in \cite{ZBC_WCNC} is referred to as the 2D-Vanilla-GNN, which satisfies the 2D-PE property. In the GNN, the pooling function is summation, the processor is linear, and the combiner is a linear function cascaded by an activation function. The hidden representation of edge $(n,k)$ in the $\ell$-th layer is obtained by the following \emph{update equation},
	\begin{equation}\label{eq:vgnn-upd-element}
		d_{nk}^{(\ell)}=\sigma\big({v}_1 d_{nk}^{(\ell-1)} + \textstyle\sum_{j=1,j\neq k}^K v_2 d_{nj}^{(\ell-1)} + \textstyle\sum_{i=1,i\neq n}^N v_3 d_{ik}^{(\ell-1)}\big),
	\end{equation}
	where $\sigma(\cdot)$ is the non-linear activation function, $v_1, v_2$ and $v_3$ are trainable weights. The second and the third terms in \eqref{eq:vgnn-upd-element} are respectively the aggregated information from neighboring edges of edge $(n,k)$ by AN$_n$ and UE$_k$. In order for the learned baseband precoding policies to be equivariant to the permutations of users and antennas, the weight $v_2$ (and also $v_3$) is identical for (i.e., shared among) all the neighboring edges. $\sigma(\cdot)$ and the weights $v_1$, $v_2$ and $v_3$ can be different among layers, but the superscript $(\ell)$ is omitted for notational simplicity.

	Denote ${\bf d}_k^{(\ell)}\triangleq[d_{1k}^{(\ell)},\cdots,d_{Nk}^{(\ell)}]^{\sf T}$ as the vector of hidden representations of the edges connected to UE$_k$, which can be seen as the updated precoding vector for UE$_k$ in the $\ell$-th layer, and can be expressed from \eqref{eq:vgnn-upd-element} as,
	\begin{eqnarray}\label{eq:vgnn-vec}
		{\mathbf{d}}_{k}^{(\ell)}=\textstyle\sigma\Big({\mathbf{V}}{\mathbf{d}}_{k}^{(\ell-1)} + {\mathbf{U}}\sum_{j=1,j\neq k}^K {\mathbf{d}}_j^{(\ell-1)}\Big),
	\end{eqnarray}
	where $\bf V$ is a square matrix whose diagonal elements are $v_1$ and non-diagonal elements are $v_3$, and ${\bf U}$ is a diagonal matrix whose diagonal elements are $v_2$. We can see that the processor is a linear element-wise function of ${\bf d}_k^{(\ell-1)}$, while the combiner is a non-linear non-element-wise function of ${\bf d}_k^{(\ell-1)}$ and a non-linear element-wise function of the aggregated vector $\sum_{j=1,j\neq k}^K {\bf d}_j^{(\ell-1)}$.
	% It can be easily proved that by stacking one or more layers after the $\ell$-th layer, the update representation is a non-linear non-element-wise function of both ${\bf d}_k^{(\ell-1)}$ and the aggregated vector.

	\subsubsection{Extended Edge-GAT}\label{sec:edge-gat}
	In \cite{Edge-GAT}, an edge-GAT was proposed for learning over homogeneous graphs, where summation is adopted as the pooling function. Since $d_{nk}^{(\ell)}$ was updated over a graph with only one type of vertices and one type of edges, a processor $\alpha(d_{nk}^{(\ell-1)}, d_{ik}^{(\ell-1)})d_{ik}^{(\ell-1)}$ was used in \cite{Edge-GAT} to extract the information from edge $(i,k)$ that is connected to edge $(n,k)$ by the $k$-th vertex, which is the neighboring edge of edge $(n,k)$ when $i\neq k$, and is the edge $(n,k)$ itself when $i=n$.  $\alpha(d_{nk}^{(\ell-1)}, d_{ik}^{(\ell-1)})$ is an attention function with trainable parameters.

	In order to learn the baseband precoding policy, we can extend the edge-GAT in \cite{Edge-GAT} for learning over the heterogeneous graph illustrated in Fig. \ref{fig:digital-precod}.
	To extract the information of edges connected with two types of vertices in the  heterogeneous graph, we use two  functions $q_{1, {\sf GAT}}(d_{nk}^{(\ell-1)}, d_{ik}^{(\ell-1)})\triangleq\alpha_1(d_{nk}^{(\ell-1)}, d_{ik}^{(\ell-1)})d_{ik}^{(\ell-1)}$ and $q_{2, {\sf GAT}}(d_{nk}^{(\ell-1)}, d_{nj}^{(\ell-1)})\triangleq\alpha_2(d_{nk}^{(\ell-1)}, d_{nj}^{(\ell-1)})d_{nj}^{(\ell-1)}$ with trainable parameters to process the information from edges $(i,k)$ and $(n,j)$, which are respectively the edges connected to edge $(n,k)$ by UE$_k$ and AN$_n$.
	Then, the hidden representation of edge $(n,k)$ in the $\ell$-th layer is updated as
	\begin{eqnarray}\label{eq:edge-gat1}
		d_{nk}^{(\ell)} \!\!&\!\!=\!\!&\!\! \textstyle\sigma\Big(\sum_{i=1}^N q_{1, {\sf GAT}}(d_{nk}^{(\ell-1)}, d_{ik}^{(\ell-1)}) +\sum_{j=1}^K q_{2, {\sf GAT}}(d_{nk}^{(\ell-1)}, d_{nj}^{(\ell-1)}) \Big)\notag\\
		\!\!&\!\!=\!\!&\!\! \textstyle\sigma\Big(q_{1, {\sf GAT}}(d_{nk}^{(\ell-1)},d_{nk}^{(\ell-1)})+q_{2, {\sf GAT}}(d_{nk}^{(\ell-1)}, d_{nk}^{(\ell-1)})+\sum_{i=1,i\neq n}^N q_{1, {\sf GAT}}(d_{nk}^{(\ell-1)}, d_{ik}^{(\ell-1)}) \notag\\
		&&\hspace{7.9cm}+\textstyle\sum_{j=1,j\neq k}^K q_{2, {\sf GAT}}(d_{nk}^{(\ell-1)}, d_{nj}^{(\ell-1)}) \Big)\notag\\
		\!\!&\!\!\triangleq\!\!&\!\! \textstyle f_{\sf GAT}\Big(d_{nk}^{(\ell-1)}, \sum_{i=1,i\neq n}^N q_{1, {\sf GAT}}(d_{nk}^{(\ell-1)}, d_{ik}^{(\ell-1)}), \textstyle\sum_{j=1,j\neq k}^K q_{2, {\sf GAT}}(d_{nk}^{(\ell-1)}, d_{nj}^{(\ell-1)}) \Big),
	\end{eqnarray}
	where $f_{\sf GAT}(\cdot)$ is a non-linear combiner.
	
	It can be seen from \eqref{eq:edge-gat1} that the combiner $f_{\sf GAT}(\cdot)$ and processors $q_{1, {\sf GAT}}(\cdot)$ and $q_{2, {\sf GAT}}(\cdot)$ are the same for different values of $n, k, i, j$, which indicates that the trainable parameters in these functions are shared among edges. With this parameter sharing scheme, it is not hard to prove that the extended edge-GAT also satisfies the 2D-PE property of the baseband precoding policy.

	From \eqref{eq:edge-gat1}, the  vector of hidden representations of all the edges connected to UE$_k$, i.e., the updated precoding vector for UE$_k$ in the $\ell$-th layer, can be expressed as,
	\begin{eqnarray}\label{eq:gat-vec}
		{\bf d}_k^{(\ell)}=\begin{bmatrix}
		d_{1k}^{(\ell)} \\ \vdots \\ d_{Nk}^{(\ell)}
		\end{bmatrix} \!\!&\!\!=\!\!&\!\!
		\begin{bmatrix}
		\textstyle f_{\sf GAT}\Big(d_{1k}^{(\ell-1)}, \sum_{i=2}^N q_{1, {\sf GAT}}(d_{1k}^{(\ell-1)}, d_{ik}^{(\ell-1)}), \textstyle\sum_{j=1,j\neq k}^K q_{2, {\sf GAT}}(d_{1k}^{(\ell-1)}, d_{1j}^{(\ell-1)}) \Big) \\ \vdots \\ f_{\sf GAT}\Big(d_{Nk}^{(\ell-1)}, \sum_{i=1}^{N-1} q_{1, {\sf GAT}}(d_{Nk}^{(\ell-1)}, d_{ik}^{(\ell-1)}), \textstyle\sum_{j=1,j\neq k}^K q_{2, {\sf GAT}}(d_{Nk}^{(\ell-1)}, d_{Nj}^{(\ell-1)})\Big)
		\end{bmatrix}\notag\\
%		\!\!&\!\!=\!\!&\!\!\textstyle\mathring{f}_{\sf GAT}\Big({\bf d}_k^{(\ell-1)}, Q_{1,{\sf GAT}}({\bf d}_k^{(\ell-1)}), \textstyle\sum_{j=1,j\neq k}^K \mathring{q}_{2, {\sf GAT}}({\bf d}_j^{(\ell-1)},{\bf d}_k^{(\ell-1)})\Big)\notag\\
		\!\!&\!\!\triangleq\!\!&\!\! \textstyle F_{\sf GAT}\Big({\bf d}_k^{(\ell-1)}, \textstyle \sum_{j=1,j\neq k}^K \mathring{q}_{2, {\sf GAT}}({\bf d}_k^{(\ell-1)},{\bf d}_j^{(\ell-1)})\Big),
	\end{eqnarray}
	where $\mathring{q}_{2, {\sf GAT}}({\bf d}_k^{(\ell-1)},{\bf d}_j^{(\ell-1)})\triangleq[q_{2, {\sf GAT}}(d_{1k}^{(\ell-1)}, d_{1j}^{(\ell-1)}),\cdots,q_{2, {\sf GAT}}(d_{Nk}^{(\ell-1)},$ $ d_{Nj}^{(\ell-1)})]^{\sf T}$, which is a non-linear element-wise function of ${\bf d}_j^{(\ell-1)}$ and ${\bf d}_k^{(\ell-1)}$ by stacking $q_{2, {\sf GAT}}(\cdot)$ from $n=1$ to $N$, and $F_{\sf GAT}(\cdot)$ is a multivariate function,
 	which is a non-linear non-element-wise function of ${\bf d}_k^{(\ell-1)}$ and a non-linear element-wise function of $\sum_{j=1,j\neq k}^K \mathring{q}_{2, {\sf GAT}}({\bf d}_k^{(\ell-1)},{\bf d}_j^{(\ell-1)})$.
	
\vspace{-2mm}\subsection{Hybrid Precoding}\label{subsec:3d-vgnn}
	The \emph{2D-Vanilla-GNN} does not perform well when learning the hybrid precoding policy due to its weak expressive power, where the channel information is lost after aggregation and combination such that the GNN cannot differentiate all channel matrices \cite{PY}. To avoid the information loss, a three-dimensional (3D)-GNN was proposed in \cite{LSJ_MultiDim_GNN_2022} where the hidden representations are updated over a hyper-graph, which contains three types of vertices (respectively corresponding to the antennas, RF chains and users) and the hyper-edges connecting to them.
	Denote the hyper-edge that connects AN$_n$, RF$_m$ (i.e., the vertex corresponding to the $m$-th RF chain), and UE$_k$ as hyper-edge $(n,m,k)$.
	% We define the hyper-graph with hyper-edges connecting to three types of vertices as \emph{3D-hyper-graph}. The definition can be extended to the case of $n$ types of vertices as \emph{$n$D-hyper-graph}.
	
	% In the $\ell$-th layer of the GNN, the representation of hyper-edge $(n,m,k)$ (denoted as $d_{nmk}^{(\ell)}$) is updated by aggregating information from the neighbored hyper-edges and then combining with $d_{nmk}^{(\ell-1)}$, where the same processing, pooling and combination functions were used as those in 2D-Vanilla-GNN.
	The hidden representations of all the hyper-edges in the $\ell$-th layer constitute a tensor ${\bf D}^{(\ell)}=[d_{nmk}^{(\ell)}]^{N\times N_s\times K}$, where $d_{nmk}^{(\ell)}$ is the hidden representation of hyper-edge $(n,m,k)$. When learning the hybrid precoding policy with the 3D-GNN in  \cite{LSJ_MultiDim_GNN_2022} containing $L$ layers, the dimension of the feature matrix ${\bf H}$ was increased to three for obtaining ${\bf D}^{(1)}$, and the dimension of ${\bf D}^{(L)}$ was decreased to two for obtaining ${\mathbf{W}}_{RF}$ and ${\mathbf{W}}_{BB}$. In the 3D-GNN, the processor, combiner and the pooling function are the same as the 2D-Vanilla-GNN, which is referred to as \emph{3D-Vanilla-GNN}. For conciseness, the update equation is no longer provided.
	
	Similar to the 2D-Vanilla-GNN, parameter sharing was introduced into the 3D-Vanilla-GNN, hence the learned policy satisfies the permutation property of the hybrid precoding policy \cite{LSJ_MultiDim_GNN_2022}.
	
%\subsection{Role of Satisfying Permutation Property on Size Generalizability}
\vspace{-2mm}\begin{remark}\label{remark:Role-PE}\emph{
			Thanks to the parameter sharing for satisfying the 2D-PE property of the baseband precoding policy, only three parameters $v_1, v_2$ and $v_3$ need to be trained in each layer of the 2D-Vanilla-GNN, regardless of the values of $K$ and $N$. Hence,
	% Since the dimension of trainable weights (e.g., $v_1, v_2$ and $v_3$ in 2D-Vanilla-GNN) is independent of $K, N$ and $N_s$,
	after being trained by the samples generated in the scenario with $K$ users and $N$ antennas, the GNN can output precoding matrices for scenarios with $K'\neq K$ users and $N'\neq N$ antennas.
	Similarly, the extended edge-GAT and the 3D-Vanilla-GNN can also output precoding matrices in different problem scales due to satisfying the permutation properties of the precoding policies. Nonetheless, the three GNNs cannot be generalized to $K$, as to be shown with the analyses and simulations later. This indicates that satisfying desired PE property is  necessary but cannot ensure a GNN for learning precoding policy to be size generalizable.
		}
	\end{remark}\vspace{-2mm}	

\vspace{-3mm}\section{Key Characteristics Enabling Size Generalizability} \label{sec: rgnn design}\vspace{-1mm}	
	In this section, we identify the key characteristics of the update equation that enable size generalizability of a GNN for learning precoding. To this end, we consider baseband and hybrid precoding problems \textbf{P1} and \textbf{P2} as two use cases, and re-express the iterative equations of the numerical algorithms for solving the two problems such that they can be regarded as the update equations of well-trained GNNs for learning the baseband and hybrid precoding policies. %By comparing them with the update equations of GNNs that are not size-generalizable, we can find the key characteristics.

	For revealing the common structure of the input-output relation of
each iteration of two numerical algorithms and facilitating the design of GNNs in the next section, we start by defining $n$D-PE property and 1D-PE function.
	
	{\bf $n$D-PE property:}
	Consider a function $\mathbf{Y}=g(\mathbf{X})$, where $\mathbf{X}$ and $\mathbf{Y}$ are order-$n$ tensors (e.g., an order-two tensor is a matrix). If the order of each dimension of $\mathbf{X}$ (e.g., the order of rows or columns of a matrix) is arbitrarily changed, the order of each dimension of $\mathbf{Y}$ is correspondingly changed while the value of $\mathbf{Y}$ keeps unchanged, then $\mathbf{Y}=g(\mathbf{X})$ is $n$D-PE to $\mathbf{X}$.
	
	For example,  a function ${\bf y}=g({\bf x})$ will satisfy 1D-PE property if ${\bf \Pi}^{\sf T}{\bf y}=g({\bf \Pi}^{\sf T}{\bf x})$, where ${\bf \Pi}$ is a permutation matrix that changes the order of elements in vectors $\bf x$ and $\bf y$, and  $n=1$. The baseband precoding policy satisfies 2D-PE property as shown in section \ref{sec: system model}, where $n=2$.

The functions that satisfy $n$D-PE property are not unique. Take $n=1$ as an	example, there are more than one functions that satisfy the 1D-PE property. By re-expressing the numerical algorithms for optimizing precoding, we will show that the input-output relation of the iterative equation of each algorithm can be expressed as a function with a special structure.
	
	\vspace{1mm}
	{\bf 1D-PE function:} A function that maps ${\bf x}\triangleq[x_1,\cdots,x_K]^{\sf T}$ to ${\bf y}\triangleq[y_1,\cdots,y_K]^{\sf T}$ in the form
	\begin{equation}\label{1D-PE function}
	y_k=f\big(x_k, \textstyle\sum_{j=1,j\neq k}^K q(x_k, x_j)\big)
	\end{equation}
satisfies the 1D-PE property because $f(\cdot)$ and $q(\cdot)$ are independent of $k$ and $\sum(\cdot)$ satisfies commutative law, which is referred to as a \emph{1D-PE function}.
	When $x_k$ and $y_k$ are replaced by tensors, the mapping in this form is still called a {1D-PE function}.

For a GNN satisfying 1D-PE property, its update equation can be expressed in the same form as in \eqref {1D-PE function} with combiner and processor consisting of trainable parameters. By borrowing the notion from GNN, $q(\cdot)$ and $f(\cdot)$ in \eqref {1D-PE function} are referred to as the processor and combiner of the 1D-PE function, respectively.

Again, the 1D-PE function is not the only function that satisfies the 1D-PE property. For example, the function that maps vector ${\bf x}$ to vector ${\bf y}$ in the form $y_k=f\big(x_k, \textstyle\sum_{j=1,j\neq k}^K q(x_j)\big)$ satisfies the 1D-PE property  \cite{WJJ_GNN_Gen} but is not referred to as a 1D-PE function.
	
\vspace{-2mm}\begin{remark}\label{remark:1d-pe-func}\emph{
			The 1D-PE function in \eqref{1D-PE function} is a \emph{non-element-wise} function, because $y_k$ depends not only on $x_k$ but also on $x_j, j\neq k$.
			When $f(\cdot)$ and $q(\cdot)$ are non-linear functions, the 1D-PE function is also a \emph{non-linear} function.
		}
	\end{remark}\vspace{-3mm}

	%	An $n$D-PE function that maps order-$n$ tensor $\bf X$ to order-$n$ tensor $\bf Y$ can be obtained by
	%	simultaneously satisfying the equivariance to the permutations of each dimension of $\bf X$ as the Vanilla-GNNs. For example, the input-output relationship of the 2D-Vanilla-GNN is a 2D-PE function. However, the 2D-Vanilla-GNN for learning baseband precoding policies cannot be well-generalized to problem scales, owing to not satisfying the three characteristics (D1)$\thicksim$(D3). In what follows, we design the RGNN with the three characteristics. The RGNN can learn $n$D-PE functions that map an order-$n$ tensor $\bf X$ to an order-$n$ tensor $\bf Y$, by recursively satisfying the equivariance to the permutations of each dimension of $\bf X$ with 1D-PE functions.
	
	We do not provide an $n$D-PE function because it is with complex form and it can be obtained by recursively satisfying the equivariance to the permutations of every dimension of $\bf X$ with 1D-PE functions as to be shown later.
		
	% Since matrix pseudo-inverse is generally required to obtain optimal baseband precoding \cite{Precod_Opt_Structure}, we consider using the GNNs to approximate ZFBF, i.e., pseudo-inverse of channel matrix. 	
	
\vspace{-4mm}\subsection{Baseband Precoding}\label{sec:analyze-gen-ability}\vspace{-1mm}
	The baseband precoding problem \textbf{P1} can be solved by the WMMSE algorithm.
\vspace{-2mm}\begin{proposition}\label{prop:wmmse-iter-eq}\emph{
		The iterative equation of the WMMSE algorithm can be re-expressed as,
		\begin{equation}\label{eq:iter-precod-wmmse}
		{\bf d}_k^{(\ell)}=f_{\sf BB}\Big({\bf d}_k^{(\ell-1)}, \textstyle\sum_{j=1,j\neq k}^K q_{\sf BB}({\bf d}_k^{(\ell-1)},{\bf d}_j^{(\ell-1)})\Big),
		\end{equation}
		where ${\bf d}_k^{(\ell)}$ is the updated vector\footnote{There is a little misuse of notations, because ${\bf d}_k^{(\ell)}$ refers to the hidden representations of all edges connected to UE$_k$ in the $\ell$-th layer of the corresponding GNNs in \eqref{eq:vgnn-vec} and \eqref{eq:gat-vec}.} for UE$_k$ in the $\ell$-th iteration given in \eqref{eq:upd-vec-bb-precod}, consisting of the updated precoding vector for UE$_k$ and other auxiliary variables, and $f_{\sf BB}(\cdot)$ and $q_{\sf BB}(\cdot)$ are respectively given in \eqref{eq:appendix-fw} and \eqref{eq:psi}.
		\begin{IEEEproof}
			See Appendix \ref{appendix:precod-wmmse}.
		\end{IEEEproof}}
	\end{proposition}\vspace{-2mm}
	
	It can be observed from \eqref{eq:iter-precod-wmmse} that the input-output relation of the iterative equation (i.e., the mapping from ${\bf d}_1^{(\ell-1)},\cdots,{\bf d}_K^{(\ell-1)}$ to ${\bf d}_k^{(\ell)}$) is a 1D-PE function. As can be seen from \eqref{eq:psi}, $q_{\sf BB}(\cdot)$ is a non-linear non-element-wise function of ${\bf d}_j^{(\ell)}$ and ${\bf d}_k^{(\ell)}$ because of involving inner product of vectors, which reflects the pair-wise correlation between the updated precoding vectors for UE$_k$ and UE$_j$. As shown in \eqref{eq:appendix-fw}, $f_{\sf BB}(\cdot)$ is also a non-linear non-element-wise function of ${\bf d}_k^{(\ell-1)}$ and the aggregated output $\sum_{j=1,j\neq k}^K q_{\sf BB}({\bf d}_k^{(\ell-1)},{\bf d}_j^{(\ell-1)})$.
	
	The iterative equation in \eqref{eq:iter-precod-wmmse} can be regarded as an update equation of a well-trained GNN (called WGNN in the sequel, where ``W'' stands for the WMMSE algorithm), in which $q_{\sf BB}(\cdot)$ and $f_{\sf BB}(\cdot)$ are respectively the corresponding processor and combiner, and ${\bf d}_k^{(\ell)}$ is the updated vector of hidden representations of the edges connected to UE$_k$ in the $\ell$-th layer.

	Since the iterative equation can converge to a local optimal solution of problem \textbf{P1} for arbitrary numbers of users, the WGNN can yield a local optimal precoding policy for every value of $K$. Moreover, we can find from Appendix \ref{appendix:precod-wmmse} that the input and output sizes of $f_{\sf BB}(\cdot)$ and $q_{\sf BB}(\cdot)$ are independent of $K$.
If a GNN is with such $K$-independent processor and combiner after well-trained for optimizing precoding under a given number of users, then the GNN can output local optimal precoding matrices for other numbers of users.
This indicates that the GNN can be generalized to $K$ owing to the following reason: \emph{the processor and the combiner do not depend on $K$ after well training.} This is the condition for a GNN learning precoding to be generalizable to $K$, which is called ``\emph{size generalization condition}'' for short in the sequel.

	% From the analysis above, we can obtain the conditions that a GNN can be well-generalized to problem scales,
%	\emph{
%		\begin{enumerate}[itemindent=2mm]
%			\item[(R1)] The WGNN can well-learn the precoding policy for each value of $K$,
%			\item[(R2)] The processing and combination functions do not depend on $K$.
%		\end{enumerate}
%	}
%	More generally, we can obtain the conditions for a GNN that learns a policy to be generalized to problem scales as follows,
%	\emph{
%		\begin{enumerate}[itemindent=5mm]
%			\item[(C2-1)] The GNN can learn the desired policy for each problem scale,
%			\item[(C2-2)] The learned processing and combination functions do not depend on problem scale.
%		\end{enumerate}
%	}
	
% \vspace{-3mm}\subsection{Key Characteristics Affecting Size Generalizability for Learning Baseband Precoding} \label{sec:analyze-gen-ability}
	%We then analyze the key characteristics that affects size generalizability.
	From \eqref{eq:vgnn-vec}, \eqref{eq:gat-vec} and \eqref{eq:iter-precod-wmmse}, we can see that %there is one difference between the combiner of the WGNN and those of the 2D-Vanilla-GNN and extended Edge-GAT. Specifically,
the combiner of the WGNN is a non-linear non-element-wise function of all the variables, while the combiners of the 2D-Vanilla-GNN and the extended edge-GAT are non-linear non-element-wise functions of ${\bf d}_k^{(\ell-1)}$ and non-linear element-wise functions of the aggregated vector. This can be seen more clearly in \eqref{eq:vgnn-vec}, where the weight ${\bf V}$ on ${\bf d}_k^{(\ell-1)}$   is a square matrix but the weight ${\bf U}$ on the aggregated output is a diagonal matrix. Nonetheless, it can be easily proven that by cascading one or more layers after the $\ell$-th layer of the extended edge-GAT and the 2D-Vanilla-GNN, the composite of combiners of multiple layers becomes a non-linear non-element-wise function of ${\bf d}_k^{(\ell-1)}$ and the aggregated vector.
	
	%We then show that the differences between the GNNs in their processors. Specifically,
By comparing \eqref{eq:vgnn-vec} and \eqref{eq:iter-precod-wmmse}, we can observe two differences between the processors of  the WGNN and the 2D-Vanilla-GNN.
	% \footnote{The term $\sum_{i=1,i\neq n}^N v_3 d_{ik}^{(\ell-1)}$ in \eqref{eq:vgnn-vec} is not in \eqref{eq:iter-precod-wmmse}, because this information has been included in $q_{\sf BB}(\cdot)$.}

		\begin{enumerate}
			\item[(D1)] A \emph{non-linear} processor $q_{\sf BB}(\cdot)$ is used in the WGNN, while a linear processor is used in the 2D-Vanilla-GNN.
			\item[(D2)] The processor $q_{\sf BB}(\cdot)$ in the WGNN \emph{reflects the pair-wise correlation between
			${\bf d}_k^{(\ell-1)}$ and ${\bf d}_j^{(\ell-1)}, j \neq k$}, while the processor of the 2D-Vanilla-GNN is only a function of ${\bf d}_j^{(\ell-1)}$.
	\end{enumerate}
	
	By comparing \eqref{eq:gat-vec} and \eqref{eq:iter-precod-wmmse}, we can observe one difference between the processors of  the WGNN and the extended edge-GAT.
		\begin{enumerate}
			\item[(D3)] A \emph{non-element-wise} processor $q_{\sf BB}(\cdot)$ is used in the WGNN, while an element-wise processor $\mathring{q}_{2,{\sf GAT}}(\cdot)$ is used in the edge-GAT.\vspace{0.1mm}
		\end{enumerate}
	
	This implies that the following {\bf three characteristics of the processor} are essential for a GNN learning precoding to be generalized to different numbers of users: \emph{non-linear, non-element-wise, and reflecting the correlation between ${\bf d}_k^{(\ell-1)}$ and ${\bf d}_j^{(\ell-1)}, j \neq k$}. To validate this, we show that a GNN satisfying the desired PE property but without satisfying the characteristics of the WGNN in (D1)$\thicksim$(D3) is not size generalizable.
	
	\subsubsection{Extended Edge-GAT} \label{sec:explain-size-gener-GAT}
	The edge-GAT with the update equation in \eqref{eq:gat-vec} is with the characteristics of the WGNN in (D1) and (D2) but without the characteristic in (D3). We explain why the edge-GAT for learning the precoding policy is not size-generalizable.

	To learn the precoding policy with the edge-GAT, one can use \eqref{eq:gat-vec} to approximate \eqref{eq:iter-precod-wmmse} that can converge to a local optimal precoding matrix after multiple layers. 	At the first glance, the approximation can be accomplished by approximating $q_{\sf BB}(\cdot)$ in \eqref{eq:psi} with the processor $\mathring{q}_{2,{\sf GAT}}(\cdot)$ and approximating $f_{\sf BB}(\cdot)$ in \eqref{eq:appendix-fw} with the combiner $F_{\sf GAT}(\cdot)$.
	Then, the size generalization condition seems able to be satisfied and the trained edge-GAT seems able to be generalized to $K$ since the two functions $q_{\sf BB}(\cdot)$ and $f_{\sf BB}(\cdot)$ do not depend on the numbers of users.
	However, since $\mathring{q}_{2,{\sf GAT}}(\cdot)$ can only approximate element-wise functions, it cannot approximate $q_{\sf BB}(\cdot)$ that is non-element-wise. Hence, \eqref{eq:appendix-fw} cannot be approximated with \eqref{eq:gat-vec} in this way.
	
	We can approximate \eqref{eq:iter-precod-wmmse} with \eqref{eq:gat-vec} in an alternative way as detailed in \cite{ModelGNN_GJ_2022}. Specifically, with the aggregated output $\sum_{j=1,j\neq k}^K \mathring{q}_{2, {\sf GAT}}({\bf d}_k^{(\ell-1)}, {\bf d}_j^{(\ell-1)})$ of the edge-GAT, one can first recover the vectors that are aggregated, i.e, ${\bf D}_{/k}^{(\ell-1)}\triangleq[{\bf d}_1^{(\ell-1)}, \cdots, {\bf d}_{k-1}^{(\ell-1)}, {\bf d}_{k+1}^{(\ell-1)},\cdots,{\bf d}_K^{(\ell-1)}]$, which contains the vectors of representations of edges connected to all the user vertices except UE$_k$. Then, the non-linear non-element-wise combiner $F_{\sf GAT}(\cdot)$\footnote{We have mentioned that the composite of the combiners in multiple layers is a non-linear non-element-wise function of all the variables. We use $F_{\sf GAT}(\cdot)$ to denote the composite of the combiners for notational simplicity, which is a little misuse of notations.} that takes ${\bf d}_k^{(\ell-1)}$ and ${\bf D}_{/k}^{(\ell-1)}$ as inputs can be applied to approximate the input-output relation of the equation in \eqref{eq:iter-precod-wmmse}, which is a composite function of $f_{\sf BB}(\cdot)$ and $q_{\sf BB}(\cdot)$.
In order for ${\bf D}_{/k}^{(\ell-1)}$ with $N(K-1)$ elements being recoverable from the aggregated output, the output dimension of $\mathring{q}_{2, \sf GAT}(\cdot)$ should be no less than $N(K-1)$ \cite{zaheer2017deep}, which changes with $K$. Also, the input of $F_{\sf GAT}(\cdot)$ consists of ${\bf D}_{/k}^{(\ell-1)}$ whose size changes with $K$.
This indicates that after being well-trained to approximate the input-output relation in \eqref{eq:iter-precod-wmmse}, the combiner and processor of the extended edge-GAT are different among problem scales, i.e., the size generalization condition is not satisfied.

	\subsubsection{Other GNNs without the Characteristics}
	We first examine two GNNs that are only with part (not all) of the three characteristics of the WGNN in (D1)$\thicksim$(D3), denoted as \emph{GNN-P} (``P'' stands for ``pair-wise correlation'') and \emph{GNN-N} (``N'' stands for ``non-linear''), respectively.
	GNN-P is with the characteristics in (D2) and (D3) but without the characteristic in (D1), where the processor is a linear function of ${\bf d}_k^{(\ell-1)}$ and ${\bf d}_j^{(\ell-1)}$.
	GNN-N is with the characteristics in (D1) and (D3) but without the characteristic in (D2), i.e., ${\bf d}_k^{(\ell-1)}$ is not a variable of a non-linear processor. For example, the GNNs proposed in \cite{GNN_Beamforming_TCOM2023,GNN_mUE_RIS_JSAC_2021_Yu} are two kinds of GNN-N, where FNNs were adopted as non-linear non-element-wise processors.
	
	Both the GNN-N and GNN-P that learn the optimal precoding policy by approximating the input-output relationship in \eqref{eq:iter-precod-wmmse} also cannot be well-generalized to problem scales. This is because
	${\bf d}_k^{(\ell-1)}$
	is not a variable of a non-linear processor in the two GNNs. As a result, the processors of both GNNs cannot approximate $q_{\sf BB}({\bf d}_k^{(\ell-1)}, {\bf d}_j^{(\ell-1)})$ in \eqref{eq:psi}. If we use the alternative way to approximate the input-output relationship in \eqref{eq:iter-precod-wmmse} as for the extended-GAT, then the vectors in ${\bf D}_{/k}^{(\ell-1)}$ should also be recoverable from the aggregated outputs of the GNN-N and GNN-P. In this way, the output of the processor and the input of the combiner are with sizes  depending on $K$. Thus, the trained GNN-N and GNN-P cannot be generalized to unseen numbers of users.

	The 2D-Vanilla-GNN is without all the three characteristics in (D1) $\sim$(D3). With a similar analysis, it is not hard to show that the 2D-Vanilla-GNN is also not size-generalizable. In Table \ref{table:size-gen}, we summarize the considered GNNs.
%, including whether each of them is with the three characteristics and whether they are size generalizable.

\vspace{-0.01mm}\begin{table}[!htb]
		\centering
		\caption{Summary of GNNs}\label{table:size-gen}\vspace{-2mm}
		\small
		\begin{tabular}{c|c|c|c|c|c}
			\hline\hline
			\multicolumn{2}{c|}{Network} & \tabincell{c}{(D1)} & \tabincell{c}{(D2)} & \tabincell{c}{(D3)} & \tabincell{c}{Size-Generalizable?}  \\ \hline
			\multicolumn{2}{c|}{WGNN}							   &	\Checkmark	&  \Checkmark & \Checkmark & \Checkmark		
			\\ \hline
			\multicolumn{2}{c|}{Extended Edge-GAT}							 &	  \Checkmark	  & \Checkmark	& \XSolidBrush	& \XSolidBrush
			\\ \hline
			\multicolumn{2}{c|}{GNN-P}							 &	  \XSolidBrush	  & \Checkmark	& \Checkmark	& \XSolidBrush		\\ \hline
			\multicolumn{2}{c|}{GNN-N}							 &	  \Checkmark	  & \XSolidBrush	& \Checkmark	& \XSolidBrush		\\ \hline
			\multicolumn{2}{c|}{2D-Vanilla-GNN}							 &	  \XSolidBrush	  & \XSolidBrush	& \XSolidBrush	& \XSolidBrush		\\ \hline
			\hline
		\end{tabular}
	\end{table}\vspace{-2mm}

If the processor of a GNN is with the characteristics of the WGNN in (D1)$\thicksim$(D3), then the combiner and processor of the GNN can respectively approximate \eqref{eq:appendix-fw} and \eqref{eq:psi}, and the size generalization condition can be satisfied.
	Otherwise, the GNN is not size generalizable.
%Hence, \emph{the three characteristics are essential for size generalizability when learning the baseband precoding policy.}
	
\vspace{-2mm}\begin{remark}\label{remark:gat-cplxty}\emph{
		For the GNNs without these characteristics (e.g., the extended edge-GAT and 2D-Vanilla-GNN), since the output size of the processor and the input size of the combiner grow with $K$, more samples are required to approximate these functions when $K$ is large.
%As a results, the sample complexity of these GNNs is high.
		% By contrast, the sample complexity of a GNN with processors satisfying all the characteristics in (D1)$\thicksim$(D3) (say, the RGNN to be designed later) does not change with $K$, because the combination and processors can be trained to respectively approximate $f_{\sf BB}(\cdot)$ and $q_{\sf BB}(\cdot)$ in \eqref{eq:iter-precod-wmmse} that are independent from $K$.
		}
	\end{remark}\vspace{-2mm}

\vspace{-4mm}\subsection{Hybrid Precoding}\vspace{-2mm}
	% It is also necessary to satisfy the three characteristics for learning size-generalizable hybrid precoding. To see this, we again re-express the iterative equation of a numerical algorithm for optimizing hybrid precoding, which can be seen as an update equation of a well-trained GNN.
	When optimizing hybrid precoding from problem \textbf{P2}, an algorithm that integrates the WMMSE algorithm and manifold optimization (MO) for solving the problem was proposed in \cite{Model_DL_HybPrec_TCOM2023}, which is referred to as \emph{WMMSE-MO} algorithm in the sequel.
	
\vspace{-2mm}\begin{proposition}\label{prop:wmmse-mo-iter-eq}\emph{
		The  iterative equation of the WMMSE-MO algorithm can be re-expressed as,
		\begin{equation}\label{eq:iter-hybrid-precod}
		{\bf D}_k^{(\ell)} = f_{\sf HB}\Big({\bf D}_k^{(\ell-1)}, \textstyle\sum_{j=1,j\neq k}^K q_{\sf HB}({\bf D}_k^{(\ell-1)}, {\bf D}_j^{(\ell-1)})\Big),
		\end{equation}
		where ${\bf D}_k^{(\ell)}$ is the updated matrix for UE$_k$ in the $\ell$-th iteration given in \eqref{eq:upd-mat-hyb-precod},  consisting of the updated values of the baseband precoding vector for UE$_k$, the analog precoding matrix, and other auxiliary variables, and $f_{\sf HB}(\cdot)$ and $q_{\sf HB}(\cdot)$ are given in \eqref{eq:f_HB} and \eqref{eq:q_HB}, respectively.
		\begin{IEEEproof}
			See Appendix \ref{appendix:hybrid-precod}.
		\end{IEEEproof}
	}
	\end{proposition}\vspace{-3mm}

	It can be observed from \eqref{eq:iter-hybrid-precod} that the input-output relation of the iterative equation (i.e., the mapping from ${\bf D}_1^{(\ell-1)},\cdots,{\bf D}_K^{(\ell-1)}$ to ${\bf D}_k^{(\ell)}$ ) is a 1D-PE function.
	
	The iterative equation can be regarded as the update equation of a well-trained GNN that learns over the hyper-graph mentioned in section \ref{subsec:3d-vgnn},
in which $q_{\sf HB}(\cdot)$ is a non-linear non-element-wise processor of ${\bf D}_k^{(\ell-1)}$ and ${\bf D}_j^{(\ell-1)}$, $f_{\sf HB}(\cdot)$ is a non-linear non-element-wise combiner of  ${\bf D}_k^{(\ell-1)}$ and the aggregated output $\sum_{j=1,j\neq k}^K q_{\sf HB}({\bf D}_k^{(\ell-1)}, {\bf D}_j^{(\ell-1)})$, and ${\bf D}_k^{(\ell)}$ is the matrix consisting of the hidden representations of all the hyper-edges connected to UE$_k$ in the $\ell$-th layer.
	
	The well-trained GNN is generalizable to the number of users, because for the iterative equation that can converge to the local optimal hybrid precoding matrix in \eqref{eq:iter-hybrid-precod}, the input and output sizes of the combiner and processor are independent of $K$ (as can be seen from Appendix \ref{appendix:hybrid-precod}), such that the size generalization condition is satisfied.
	
	As shown in \eqref{eq:iter-hybrid-precod}, the well-trained GNN is with the three characteristics of the WGNN, i.e., $f_{\sf HB}(\cdot)$ is a non-linear non-element-wise processor that reflects the correlation between ${\bf D}_k^{(\ell-1)}$ and ${\bf D}_j^{(\ell-1)}$.\footnote{The well-trained GNN reflects the correlation between ${\bf D}_k^{(\ell-1)}$ and ${\bf D}_j^{(\ell-1)}$ instead of ${\bf d}_k^{(\ell-1)}$ and ${\bf d}_j^{(\ell-1)}$ as stated in the characteristic in (D2) because the GNN is learned over a hyper-graph instead of the graph shown in Fig. \ref{fig:digital-precod}.} According to previous analyses, in order for a GNN to learn a hybrid precoding policy with size generalizability by approximating \eqref{eq:iter-hybrid-precod}, the GNN should also be with the characteristics.

\vspace{-3mm}\section{Design of Size Generalizable GNNs}\label{sec:rgnn-design}\vspace{-1mm}
	In this section, we design the update equations of the GNNs that are with the three characteristics meanwhile satisfying the desired permutation properties. We first propose a size generalizable GNN for learning the baseband precoding policy, then extend the design for learning hybrid precoding policy, and finally show how to design the GNN for learning over hyper-graph with $n$ types of vertices and for learning other wireless policies. Since the proposed GNN satisfies the high-dimensional permutation property in a recursive manner, it is referred to as  \emph{Recursive GNN (RGNN)}.
	
\vspace{-5mm}\subsection{RGNN for Learning Baseband Precoding}\label{subsec:2d-rgnn}\vspace{-1mm}
	Denote the hidden representation of the RGNN in the $\ell$-th layer as ${\bf D}^{(\ell)}=[{\bf d}_1^{(\ell)},\cdots,{\bf d}_K^{(\ell)}]\in{\mathbb R}^{N\times K}$, which is obtained from ${\bf D}^{(\ell-1)}$ with an update equation that should be equivariant to the permutations of the antenna and user vertices (i.e., equivariant to the permutations of rows and columns of ${\bf D}^{(\ell-1)}$).
To design a GNN with the three characteristics for learning baseband precoding policies from more problems rather than only from \textbf{P1}, we consider the following update equation that has the same structure as in \eqref{eq:iter-precod-wmmse},
	\begin{equation}\label{eq:sgnn-1}
		{\mathbf{d}}_k^{(\ell)} = \textstyle f_{\mathsf{R}}\Big({\mathbf{d}}_k^{(\ell-1)}, \sum_{j=1, j\neq k}^K q_{\mathsf{R}}({\mathbf{d}}_k^{(\ell-1)}, {\mathbf{d}}_j^{(\ell-1)})\Big),
	\end{equation}
	where $q_{\mathsf{R}}(\cdot)$ is a non-linear non-element-wise processor of ${\mathbf{d}}_k^{(\ell-1)}$ and ${\mathbf{d}}_j^{(\ell-1)}$, and $f_{\mathsf{R}}(\cdot)$ is a non-linear non-element-wise combiner of ${\mathbf{d}}_k^{(\ell-1)}$ and the aggregated output $\sum_{j=1, j\neq k}^K q_{\mathsf{R}}({\mathbf{d}}_k^{(\ell-1)}, {\mathbf{d}}_j^{(\ell-1)})$. However, different from  $f_{\sf BB}(\cdot)$ and $q_{\sf BB}(\cdot)$ in \eqref{eq:iter-precod-wmmse} with the given forms in \eqref{eq:appendix-fw} and \eqref{eq:psi}, $f_{\mathsf{R}}(\cdot)$ and $q_{\mathsf{R}}(\cdot)$ can be learned with trainable weights, where $q_{\mathsf{R}}(\cdot)$ is able to learn the correlation between ${\mathbf{d}}_k^{(\ell-1)}$ and  ${\mathbf{d}}_j^{(\ell-1)}$.
	% The update of ${\bf d}_1^{(\ell)}$ with RGNN is illustrated in Fig. \ref{fig:fig-2d-rgnn}. It can be seen that each user, all the antennas and the edges connecting them are seen together as a ``hyper-vertex'' (shown with the shadowed circle), and the representation of the $k$-th hyper-vertex is ${\bf d}_k^{(\ell)}$. To update ${\bf d}_1^{(\ell)}$, the RGNN first extract information from the other two hyper-vertices with $q_{\sf R}(\cdot)$ and aggregate them with pooling function $\Sigma(\cdot)$. Then, the aggregated information is combined with  ${\bf d}_1^{(\ell-1)}$ by the combiner $f_{\sf R}(\cdot)$.
%As to be shown in section \ref{sec:simulation-results}, the GNN can learn precoding policies other than SE-maximal precoding, such as EE-maximal precoding.
	
	%It can be proved that the update equation \eqref{eq:sgnn-1} is equivariant to the permutations of users, because $f(\cdot)$ and $q(\cdot)$ are the same for different $k$ (i.e., different users), and the pooling function $\sum(\cdot)$ satisfies commutative law.
	%\begin{definition}\emph{
	%	A function that maps ${\bf x}=[x_1,\cdots,x_K]^{\sf T}$ to ${\bf y}=[y_1,\cdots,y_K]^{\sf T}$ with the form of \eqref{eq:sgnn-1}, i.e., $y_k=f\big(x_k, \sum_{j=1,j\neq k}^K q(x_k, x_j)\big)$, is defined as \emph{1D-PE function}. Here $x_k$ and $y_k$ can also be vectors, matrices or tensors.
	%	}
	%\end{definition}

	The input-output relation of the update equation in \eqref{eq:sgnn-1} is with the same structure as in \eqref{1D-PE function}, which is a 1D-PE function that is equivariant to the permutations of the users.
%columns of ${\bf D}^{(\ell-1)}$.
	
	To further ensure the equivariance to the permutations of antennas
%rows of ${\bf D}^{(\ell-1)}$ (i.e., permutations of elements in ${\bf d}_k^{(\ell-1)}, \forall k$)
and to ensure that $f_{\mathsf{R}}(\cdot)$ and $q_{\mathsf{R}}(\cdot)$ are non-element-wise functions, $f_{\mathsf{R}}(\cdot)$ and $q_{\mathsf{R}}(\cdot)$ can be designed as 1D-PE functions. To ensure the non-linearity of the two functions, either the  combiners or the processors of these two 1D-PE functions should be non-linear.
%We can simply let both the combiners and processors of the two 1D-PE functions be non-linear.

Take $f_{\mathsf{R}}(\cdot)$ that maps the vector ${\bf x}\triangleq [{\mathbf{d}}_k^{(\ell-1)}, \sum_{j=1,j\neq k}^K q_{\mathsf{R}}({\mathbf{d}}_k^{(\ell-1)}, {\mathbf{d}}_j^{(\ell-1)})]$ to the vector ${\bf y}\triangleq {\mathbf{d}}_k^{(\ell)}$ as an example, which can be designed as a 1D-PE function  as,
	\begin{equation}\label{eq:1d-pe}
		y_n = \psi_f\Big(x_n, \textstyle\sum_{i=1,i\neq n}^N \xi_f(x_n, x_i)\Big).
	\end{equation}
%	where $x_n$ and $y_n$ are the $n$-th element of ${\bf x}$ and ${\bf y}$, respectively.
Similarly, $q_{\mathsf{R}}(\cdot)$ can be designed as a 1D-PE function that is with non-linear processor and non-linear combiner, respectively denoted as $\xi_q(\cdot)$ and $\psi_q(\cdot)$.
	The combiners $\psi_f(\cdot)$, $\xi_f(\cdot)$ and processors $\xi_q(\cdot)$ and $\psi_q(\cdot)$ can be simply adopted as FNNs.
	
\begin{figure}[!htb]
		\centering
		\includegraphics[width=\linewidth]{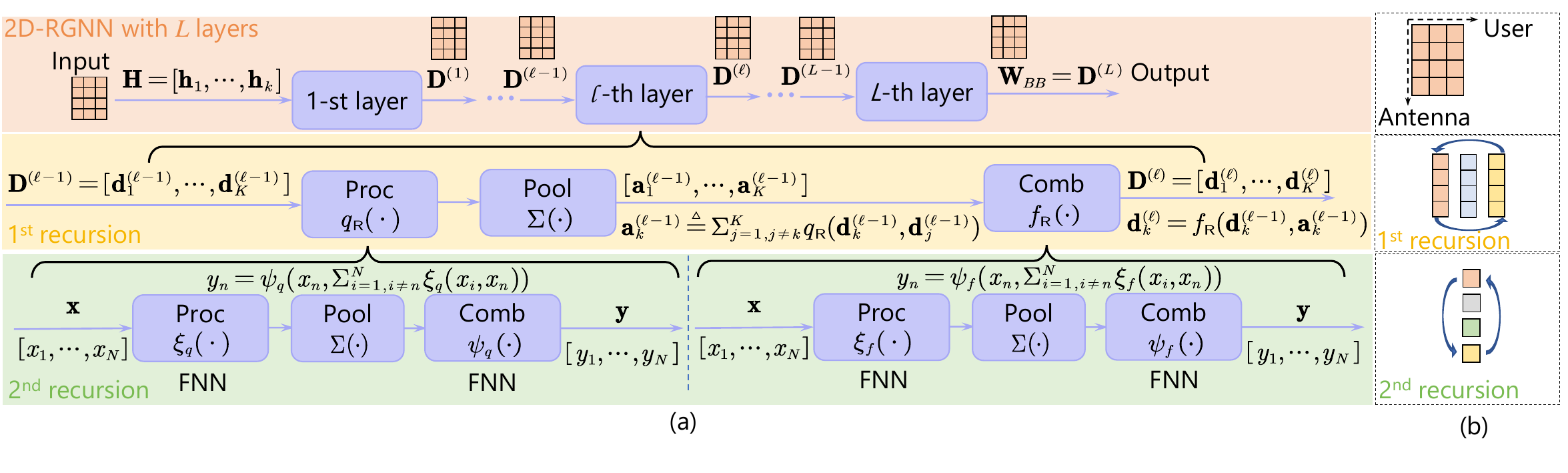}\vspace{-3mm}
		\caption{Illustration of 2D-RGNN with $L$ layers, where ``Proc'', ``Pool'' and ``Comb'' respectively refer to processor, pooling function and combiner.}
		\label{fig:fig-2d-rgnn}
	\end{figure}

	We refer to the designed GNN with the update equation in \eqref{eq:sgnn-1} where the processor and combiner are 1D-PE functions in \eqref{eq:1d-pe} as 2D-RGNN, which satisfies the 2D-PE property recursively. The input and output of the 2D-RGNN are respectively the channel matrix ${\bf H}$ and the learned baseband precoding matrix ${\bf W}_{BB}$. The output of each layer is a matrix. Each layer (say the $\ell$-th layer) consists of two recursions for satisfying the equivariance to the permutations of columns and rows in ${\bf D}^{(\ell-1)}$, respectively. The first recursion includes one 1D-PE function, and the second recursion includes two 1D-PE functions, where each 1D-PE function is composed of processor, pooling, and combiner. The architecture of the 2D-RGNN is shown in Fig. \ref{fig:fig-2d-rgnn}.
	% The output of each layer is an matrix. In the first recursion, the update equation is designed as a 1D-PE function to guarantee the equivariance to the permutations of columns of ${\bf D}^{(\ell-1)}$. In the second recursion, the processor and combiner are further designed as 1D-PE functions to guarantee the equivariance to the permutations of rows of ${\bf D}^{(\ell-1)}$.
	
\vspace{-3mm}\subsection{RGNN for Learning Hybrid Precoding} \label{subsec:3d-rgnn}
	In each layer of the RGNN (say the $\ell$-th layer), the representations of all the hyper-edges are updated, which form an order-three tensor ${\bf D}^{(\ell)}\in{\mathbb R}^{N\times N_s \times K}$ with antenna-, RF chain- and user-dimensions.
	It was proved in \cite{LSJ_MultiDim_GNN_2022} that the three-set permutation property of the optimal policy can be satisfied if (i) the dimensions of input and output of  a GNN are respectively increased and decreased with the method in \cite{LSJ_MultiDim_GNN_2022}, and
	(ii) the update equation of the GNN for updating ${\bf D}^{(\ell)}$ from ${\bf D}^{(\ell-1)}$ satisfies 3D-PE property such that it is equivariant to the permutations of user, antenna and RF chain vertices (i.e., equivariant to the permutations in the three dimensions of ${\bf D}^{(\ell-1)}$).  We next show how to design the update equation of the RGNN to satisfy condition (ii).
	% It can be proved that if ${\bf D}^{(\ell)}$ is 3D-PE to ${\bf D}^{(\ell-1)}$, then the learned hybrid precoding policy is equivariant to the permutations of users, antennas and RF chains, which satisfies the PE property of the optimal policy.
	
	In order for the RGNN to be with the three characteristics, the hidden representations of all the hyper-edges connected to UE$_k$ and UE$_j$ in the $(\ell-1)$-th layer should be variables of a non-linear and non-element-wise processor. The update equation of the hidden representations of the hyper-edges connected to UE$_k$ in the $\ell$-th layer can be designed as follows,
	\begin{equation}\label{eq: hybrid upd}
		{\mathbf{D}}_k^{(\ell)}=\textstyle f_{\mathsf{R}}\Big({\mathbf{D}}_k^{(\ell-1)}, \sum_{j=1,j\neq k}^K q_{\mathsf{R}}({\mathbf{D}}_k^{(\ell-1)}, {\mathbf{D}}_j^{(\ell-1)})\Big),
	\end{equation}
	where ${\bf D}_k^{(\ell)}\in{\mathbb R}^{N\times N_s}$
is the $k$-th slice of the tensor ${\bf D}^{(\ell)} \in{\mathbb R}^{N\times N_s \times K} $ along its third dimension, $q_{\mathsf{R}}(\cdot)$ is a non-linear non-element-wise processor of ${\mathbf{D}}_k^{(\ell-1)}$ and ${\mathbf{D}}_j^{(\ell-1)}$, and $f_{\mathsf{R}}(\cdot)$ is a non-linear non-element-wise combiner of ${\mathbf{D}}_k^{(\ell-1)}$ and the aggregated output $\sum_{j=1, j\neq k}^K q_{\mathsf{R}}({\mathbf{D}}_k^{(\ell-1)}, {\mathbf{D}}_j^{(\ell-1)})$. Different from $f_{\sf HB}(\cdot)$ and $q_{\sf HB}(\cdot)$ in \eqref{eq:iter-hybrid-precod} with the given forms in \eqref{eq:f_HB} and \eqref{eq:q_HB}, $f_{\mathsf{R}}(\cdot)$ and $q_{\mathsf{R}}(\cdot)$ can be learned with trainable weights.
	
	% The update of ${\bf D}_1^{(\ell)}$ with RGNN is illustrated in Fig. \ref{fig:fig-3d-rgnn}. It can be seen that each user, all the antennas, all the RF chains and the hyper-edges connecting them are seen together as a ``hyper-vertex'' (shown with the shadowed circle), and the representation of the $k$-th hyper-vertex in the $\ell$-th layer is ${\bf D}_k^{(\ell)}$. To update ${\bf D}_1^{(\ell)}$, the RGNN first extract information from the other two hyper-vertices with $q_{\sf R}(\cdot)$ and aggregate them with pooling function $\Sigma(\cdot)$. Then, the aggregated information is combined with  ${\bf D}_1^{(\ell-1)}$ by the combiner $f_{\sf R}(\cdot)$.
	
	Again, the update equation is with the same form as  in \eqref{1D-PE function} where $x_k$ and $y_k$ in \eqref{1D-PE function} are replaced by matrices ${\bf D}_k^{(\ell)}$ and ${\bf D}_k^{(\ell-1)}$. The input-output relation of the update equation is a 1D-PE function, such that it is equivariant to the permutations in the user-dimension of ${\bf D}^{(\ell-1)}$.
	
	In order for the GNN to be equivariant to the permutations of the RF chain- and antenna-dimensions in ${\bf D}^{(\ell-1)}$ (i.e., the permutations of columns and rows in ${\bf D}_k^{(\ell-1)}, \forall k$), $f_{\mathsf{R}}(\cdot)$ and $q_{\mathsf{R}}(\cdot)$ should satisfy 2D-PE property.
Different from the 2D-RGNN, $f_{\sf R}(\cdot)$ and $q_{\sf R}(\cdot)$ can be designed in different ways to satisfy the 2D-PE property. We take $q_{\sf R}(\cdot)$ that maps ${\bf X}\triangleq[{\mathbf{D}}_k^{(\ell-1)}, {\mathbf{D}}_j^{(\ell-1)}]$ to ${\bf Y}\triangleq q_{\mathsf{R}}({\mathbf{D}}_k^{(\ell-1)}, {\mathbf{D}}_j^{(\ell-1)})$ as an example.
	\begin{enumerate}
		\item $q_{\sf R}(\cdot)$ can be designed as a 2D-Vanilla-GNN layer in section \ref{sec:2d-vgnn}, where the $m$-th column of ${\bf Y}$ (denoted as ${\bf y}_m$) is updated with non-linear combiner and linear processor as,
		\begin{equation}\label{eq:vgnn}
			{\bf y}_m = \sigma\big({\bf V}{\bf x}_m + {\bf U}\textstyle\sum_{i=1,i\neq m}^{N} {\bf x}_i\big),
		\end{equation}
		where ${\bf x}_m$ is the $m$-th column of ${\bf X}$.
		\item $q_{\sf R}(\cdot)$ can designed as an extended edge-GAT layer in section \ref{sec:edge-gat}, where ${\bf y}_m$ is updated with non-linear combiner and non-linear processor as,
		\begin{equation}\label{eq:edge-gat}
			{\bf y}_m = \textstyle F_{\sf GAT}\Big({\bf x}_m, \textstyle \sum_{i=1,i\neq m}^{N} \mathring{q}_{2, {\sf GAT}}({\bf x}_i,{\bf x}_m)\Big),
		\end{equation}
where $F_{\sf GAT}(\cdot)$ and $\mathring{q}_{2, {\sf GAT}}(\cdot)$ were defined in \eqref{eq:gat-vec}.
		\item $q_{\sf R}(\cdot)$ can be designed as a 2D-RGNN layer in section \ref{subsec:2d-rgnn}. Specifically, ${\bf y}_m$ is obtained by a composite function of $\psi_f(\cdot)$ and $\xi_f(\cdot)$ as,
		\begin{equation}\label{eq:2d-rgnn}
			{\mathbf{y}}_{m}=\psi_f\Big({\mathbf{x}}_{m}, \textstyle \sum_{i=1,i\neq m}^{N} \xi_f({\mathbf{x}}_{m}, {\mathbf{x}}_{i})\Big),
		\end{equation}
		which is a 1D-PE function to ensure the equivariance to the permutations in the antenna-dimension of ${\bf D}^{(\ell-1)}$.
		% with processing and combination functions being $\xi_f(\cdot)$ and $\phi_f(\cdot)$, respectively.
		To further guarantee the equivariance to the permutations in RF chain-dimension of ${\bf D}^{(\ell-1)}$, $\psi_f(\cdot)$ and $\xi_f(\cdot)$ can further be designed as 1D-PE functions, which can adopt FNNs as the non-linear processor and combiner.
	\end{enumerate}

	As analyzed in section \ref{sec:2d-vgnn} and section \ref{sec:edge-gat}, $q_{\sf R}(\cdot)$ designed in the first or the second way is a non-linear non-element-wise function of ${\bf x}_m$ but is a non-linear element-wise function of the aggregated output.
% (i.e., $\sum_{i=1,i\neq m}^{N} {\bf x}_i$ for the first way and $\sum_{i=1,i\neq m}^{N} \mathring{q}_{2, {\sf GAT}}({\bf x}_i,{\bf x}_m)$ for the second way).
 This indicates that the designed $q_{\sf R}(\cdot)$ is not a non-element-wise function of all the variables, hence is not with the characteristic in (D3). By contrast, since $\psi_f(\cdot)$ is a 1D-PE function with non-linear processor and combiner, the designed $q_{\sf R}(\cdot)$ in the third way is a non-linear non-element-wise function of all the variables according to Remark \ref{remark:1d-pe-func}
% . Hence, $q_{\sf R}(\cdot)$ designed in the third way is a non-linear non-element-wise function of all the variables,
 such that the update equation in \eqref{eq: hybrid upd} is with all the three characteristics in (D1)$\sim$(D3).
	
\begin{figure}[!htb]
		\centering
		\includegraphics[width=\linewidth]{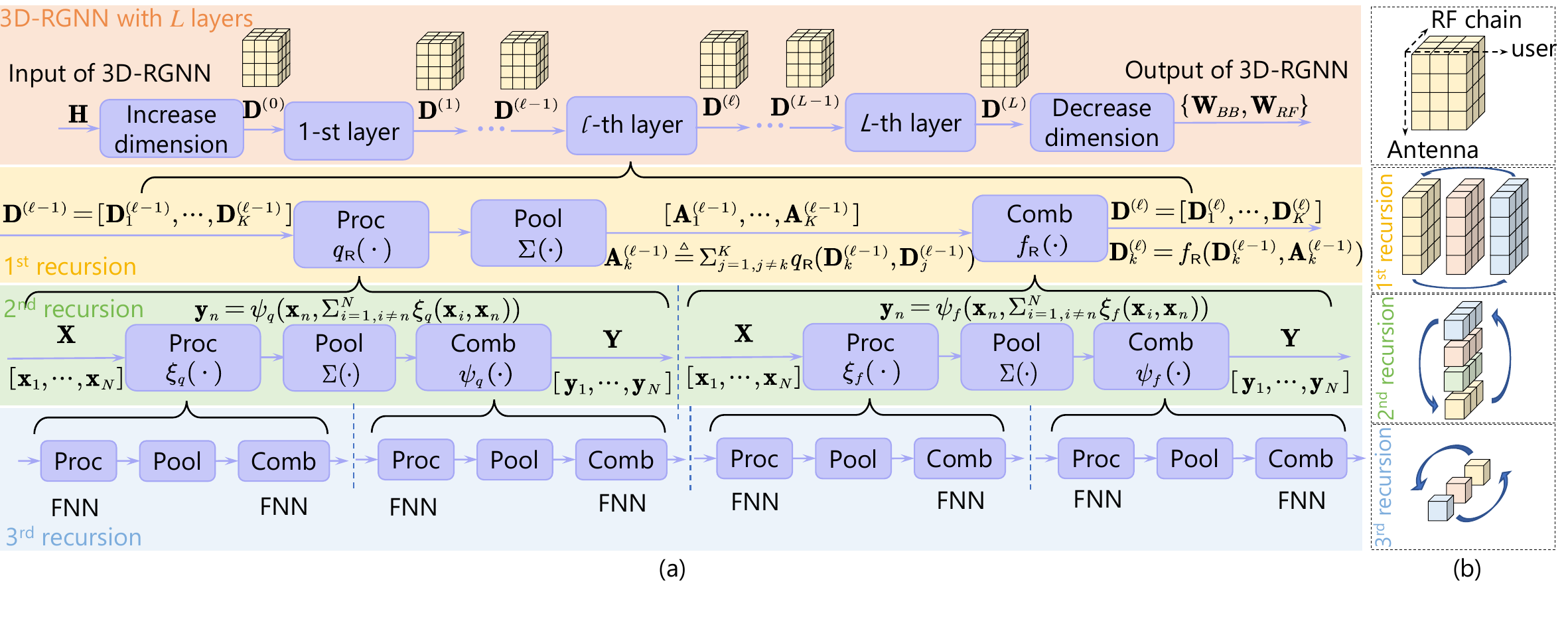}\vspace{-4mm}
		\caption{Illustration of 3D-RGNN with $L$ layers. }
		\label{fig:fig-3d-rgnn}
	\end{figure}

	We refer to the GNN with the update equation in \eqref{eq: hybrid upd} in which the processor and combiner are in \eqref{eq:2d-rgnn} as 3D-RGNN, which satisfies the 3D-PE property recursively.
	The input and output of the 3D-RGNN are respectively $\bf H$ and $\{{\bf W}_{BB}, {\bf W}_{RF}\}$, whose dimensions are respectively increased to obtain ${\bf D}^{(1)}$ and decreased from ${\bf D}^{(L)}$ with the method in \cite{LSJ_MultiDim_GNN_2022}. The output of each layer is an order-three tensor. Each layer (say the $\ell$-th layer) consists of three recursions for satisfying the equivariance to the permutations in the user-, antenna- and RF chain-dimension in ${\bf D}^{(\ell-1)}$, respectively, and the $i$-th recursion includes $2^{i-1}$ 1D-PE functions. The architecture of the 3D-RGNN is illustrated in Fig. \ref{fig:fig-3d-rgnn}.
	% The output of each layer is an order-three tensor with user-, RF chain- and antenna-dimension. In the first recursion, the update equation is designed as a 1D-PE function to guarantee the equivariance to the permutations of user-dimension of ${\bf D}^{(\ell-1)}$. In the second recursion, the processor and combiner are further designed as 1D-PE functions to guarantee the equivariance to the permutations of antenna-dimension of ${\bf D}^{(\ell-1)}$. In the third recursion, the process is repeated to guarantee the equivariance to the permutations of RF chain-dimension of ${\bf D}^{(\ell-1)}$.
	
\vspace{-5mm}\subsection{Extension to $n$D-RGNN and Other Wireless Policies}\vspace{-1mm}
\subsubsection{Extension to $n$D-RGNN}\vspace{-0.1mm}
	When learning over a hyper-graph with $n$ types of vertices and hyper-edges connecting the vertices, the output of the $\ell$-th layer is an order-$n$ tensor ${\bf D}^{(\ell)}\in{\mathbb R}^{K_1\times\cdots\times K_n}$.
	Each dimension of ${\bf D}^{(\ell)}$ corresponds to one type of vertices  \cite{LSJ_MultiDim_GNN_2022}.
	% Taking learning hybrid precoding over the hyper-graph mentioned in section \ref{subsec:3d-vgnn} as an example, ${\bf D}^{(\ell)}$ is an order-three tensor with user-dimension, antenna-dimension and RF chain-dimension, as shown in Fig. \ref{fig:fig-3d-rgnn}.
	The update equation of the RGNN (denoted as $n$D-RGNN) in the $\ell$-th layer should be $n$D-PE to ${\bf D}^{(\ell-1)}$, i.e., equivariant to the permutations of every dimension of ${\bf D}^{(\ell-1)}$, in order for learning precoding policies with size-generalizability.
	
	Analogous to the 3D-RGNN, the update equation of $n$D-RGNN in the $\ell$-th layer to update ${\bf D}^{(\ell)}$ with ${\bf D}^{(\ell-1)}$ can be designed in a recursive manner, which is shown in Procedure \ref{algo: train drnn}. Specifically, the user-dimension is firstly selected, and the update equation can be designed as \eqref{eq:1d-pe-func} such that ${\bf D}^{(\ell)}$ is equivariant to the permutations of the user-dimension of ${\bf D}^{(\ell-1)}$, similar to \eqref{eq:sgnn-1} for the 2D-RGNN and \eqref{eq: hybrid upd} for the 3D-RGNN.
	% Here we firstly select the user-dimension such that the processor reflects correlation between ${\bf D}_k^{(\ell)}$ and ${\bf D}_j^{(\ell)}$, which are the representations of all the hyper-edges connected to UE$_k$ and UE$_j$, respectively.
	By selecting the user-dimension firstly, ${\bf X}_k$ and ${\bf X}_j$ in the processor in \eqref{eq:1d-pe-func} are respectively ${\bf D}_k^{(\ell-1)}$ and ${\bf D}_j^{(\ell-1)}$. Then,
	the processor can model the pair-wise correlation between ${\bf D}_k^{(\ell-1)}$ and ${\bf D}_j^{(\ell-1)}$, which is with the characteristic in (D2) in section \ref{sec:analyze-gen-ability}.
	The processor should also be non-linear and non-element-wise according to the characteristics in (D1) and (D3), and
	the two functions should be equivariant to the permutations in the remaining $n-1$ dimensions of ${\bf D}^{(\ell-1)}$. To ensure these, another dimension is selected from the remaining $n-1$ dimensions, say the $i'$-th dimension, and the two functions are again designed as  1D-PE functions such that the two functions are equivariant to the permutations in the $i'$-th dimension of ${\bf D}^{(\ell-1)}$. The process continues repeatedly until the equivariance to the permutations in all the $n$ dimensions of ${\bf D}^{(\ell-1)}$ is satisfied.

\vspace{-0.1mm}\begin{algorithm}
		\caption{Designing the update equation of $n$D-RGNN}\label{algo: train drnn}
		\begin{algorithmic}[1]
			\State ${\bf Y}\leftarrow{\bf D}^{(\ell)}, {\bf X}\leftarrow {\bf D}^{(\ell-1)}, {\cal M}=\{1,\cdots,n\}$ is the set of all the dimensions of ${\bf D}^{(\ell)}$.
			\While {${\mathcal{M}}\neq \emptyset$}
			\Statex ~~~~\emph{Goal: Design function ${\bf Y}=F({\bf X})$  that is $|{\cal M}|$D-PE to ${\bf X}$.}
			\State Select a dimension from ${\cal M}$, e.g., the $i$-th dimension (the firstly selected dimension is user-dimension)
			\State The function that is equivariant to the permutations of the $i$-th dimension is designed as
			\Statex \vspace{-3mm}\begin{equation}{\mathbf{Y}}_{k}=\textstyle f_{\mathsf{R}i}\big({\mathbf{X}}_{k}, \sum_{j=1, j\neq k}^{K_i} q_{\mathsf{R}i}({\mathbf{X}}_{k}, {\mathbf{X}}_j)\big),\label{eq:1d-pe-func}\end{equation}
			\Statex ~~~~where ${\bf X}_k$ and ${\bf Y}_k$ are respectively the $k$-th slice of the tensors ${\bf X}$ and ${\bf Y}$ along their $i$-th dimension
			\State $f_{\mathsf{R}i}(\cdot)$ and $q_{\mathsf{R}i}(\cdot)$ need to be designed to satisfy $(|{\mathcal{M}}|\!-\!1)$D-PE property
			\State Remove the $i$-th dimension from $\cal M$
			\EndWhile
		\end{algorithmic}
	\end{algorithm}

	The $n$D-RGNN can be used to learn precoding policies over hyper-graphs containing $n$ types of vertices, such as learning hybrid precoding in a wide-band MISO system over a hyper-graph containing four types of vertices corresponding to users, RF chains, antennas and subcarriers.

\vspace{-0.2mm}\subsubsection{Extension to Other Problems}\vspace{-0.1mm}	
%\vspace{-2mm}\begin{remark}\emph{
Different from the deep unfolding networks designed by mimicking the iteration equations of numerical algorithms \cite{DeepUnfold_WMMSE_Arindam_2020, DeepUnfold_WMMSE_TWC_2021}, the design of the RGNN is inspired by re-expressing the iterative equations of numerical algorithms in the form of the update equations of GNNs, which do not depend on mathematical models. Hence, the RGNN can be flexibly applied to learn precoding policies from problems with different objective functions and constraints, learn  precoding policies in different systems such as narrow- and wide-band multi-user MISO systems, and  learn end-to-end precoding policies (i.e., the mapping from uplink pilots to downlink precoding) from different problems.

By analyzing the iterative equations of the numerical algorithms for problems other than precoding and expressing them in the form of update equations, the same method can be used to design size-generalizable GNNs for other kinds of wireless policies.

	\subsection{Sample Complexity and Inference Complexity}
	In the following, we analyze the sample complexity and inference complexity of the RGNN, which are respectively measured by the number of samples for training and the number of floating-point operations (FLOPs) for inference.
	
	\subsubsection{Sample Complexity versus $K$}\label{sec:rgnn-complexity}
	We can see from the update equation of the RGNN in \eqref{eq:sgnn-1} or \eqref{eq: hybrid upd} that only two functions $f_{\sf R}(\cdot)$ and $q_{\sf R}(\cdot)$ need to be learned in each layer, which are identical for all the users. Then, when training the RGNN with a sample generated in a scenario with $K$ users, the sample can be regarded as $K$ ``equivalent samples'' each used for learning the two functions, as explained in the sequel by taking the 2D-RGNN as an example.
	
	For the simplicity of analysis, we assume that the RGNN is only with one layer. Then, the output of the 2D-RGNN for learning the baseband precoding policy can be written as ${\bf w}_{BB_k}=f_{\sf R}({\bf h}_k, \sum_{j=1,j\neq k}^K q_{\sf R}({\bf h}_k, {\bf h}_j)), k=1,\cdots,K$. We consider unsupervised learning, hence each sample only contains the input of the GNN (i.e.,  ${\bf H}$) and does not contain the expected output. Denote $\mathcal L$ as the loss function, which is the negative sum rate averaged over all the training samples when learning the policy from problem \textbf{P1}. The gradient for finding the free parameters in $f_{\sf R}(\cdot)$ (denoted as $\bm{\theta}_f$) with one sample can be obtained as,
	\begin{equation}\label{eq:loss-func}
		 \nabla_{\bm\theta_f}=\sum_{k=1}^K\frac{\partial{\mathcal L}}{\partial {\bf w}_{BB_k}}\frac{\partial{\bf w}_{BB_k}}{\partial \bm{\theta}_f}.
	\end{equation}

We can see from \eqref{eq:loss-func} that $\nabla_{\bm\theta_f}$ is a summation of $K$ terms, which can be seen as the summation of the gradient with respect to $\bm{\theta}_f$ over $K$ ``equivalent samples''.
	In other words, the sample ${\bf H}$ is used for $K$ times for optimizing $\bm{\theta}_f$.
	The free parameters in $f_{\sf R}(\cdot)$ can be better trained with more equivalent samples.
With a similar analysis, we can also see that one sample is used as $K$ equivalent samples to train $q_{\sf R}(\cdot)$.
	
	If $N_{f_{\sf R}}$ and $N_{q_{\sf R}}$ ``equivalent samples'' are required to respectively learn $f_{\sf R}(\cdot)$ and $q_{\sf R}(\cdot)$ such that the RGNN can achieve an expected performance, we only need $\max \{\lceil\frac{N_{f_{\sf R}}}{K}\rceil, \lceil\frac{N_{q_{\sf R}}}{K}\rceil\}$ samples generated in the scenario with $K$ users to respectively well-learn the two functions. This indicates that the number of samples for training the RGNN decreases with $K$, i.e., \emph{as the problem scale grows, the sample complexity for training the RGNN reduces.}
	
	This conclusion is contradicted with the trend  in Remark \ref{remark:gat-cplxty} for the extended edge-GAT, GNN-N, GNN-P and 2D-Vanilla-GNN, which are also with parameter-sharing among users. The reason is as follows. For a RGNN that is with the three characteristics in (D1)$\sim$(D3), the processor and combiner for well-learning precoding policy are identical among problem scales, hence the number of ``equivalent samples'' required for well-learning the two functions  are identical for different values of $K$. By contrast, for the other four GNNs,  the processors and combiners for well-learning precoding policies are different among $K$ as analyzed in section \ref{sec:analyze-gen-ability}. Hence, $N_{f_{\sf R}}$ and $N_{q_{\sf R}}$ are also different among $K$, and the inverse relationship between the sample complexity and $K$ no longer holds.

	\subsubsection{Number of FLOPs of Inference}
For a $N\times K$ matrix $\bf A$ and a $K\times M$ matrix $\bf B$, $MNK$ additions and multiplications are required for computing ${\bf AB}$, hence the number of FLOPs for matrix multiplication is ${\cal O}(MNK)$. For a FNN with each layer containing $M$ neurons, a multiplication of a $M\times M$ weight matrix and a $M\times 1$ vector is required in each layer, and hence the number of FLOPs of the FNN with $L_F$ layers is ${\cal O}(L_FM^2)$.
	
	For a 1D-RGNN learning over a graph with $N$ vertices, the hidden representation of the $n$-th vertex can be updated as $d_n^{(\ell)}=f_{\sf R}(d_n^{(\ell-1)}, \sum_{i=1,i\neq n}^N q_{\sf R}(d_n^{(\ell-1)}, d_i^{(\ell-1)}))$, where $f_{\sf R}(\cdot)$ and $q_{\sf R}(\cdot)$ are FNNs each with $L_F$ layers. Each  hidden representation can be a vector (say $\mathbf{d}_n^{(\ell)}$) instead of a scalar $d_n^{(\ell)}$, and we denote the number of elements in the representation vector of the $\ell$-th layer as $J_{\ell}$. For the simplicity of analysis, we assume that the number of neurons in each layer of the FNN is identical. Then, the number of FLOPs for computing $f_{\sf R}(\cdot)$ and $q_{\sf R}(\cdot)$ is ${\cal O}(L_FJ_{\ell}^2)$. Since $q_{\sf R}(\mathbf{d}_n^{(\ell-1)}, \mathbf{d}_i^{(\ell-1)}), i=1,\cdots,N$ need to be computed to obtain $\mathbf{d}_n^{(\ell)}$, the number of FLOPs for computing $\mathbf{d}_n^{(\ell)}$ is ${\cal O}(L_FNJ_{\ell}^2)$, and the number of FLOPs of the $\ell$-th 1D-RGNN layer for computing the  hidden representations of all the $N$ vertices is ${\cal O}(L_F N^2 J_{\ell}^2)$.
	
	For a 2D-RGNN learning over the graph in Fig. \ref{fig:digital-precod}, the update equation is provided in \eqref{eq:sgnn-1}, where $f_{\sf R}(\cdot)$ and $q_{\sf R}(\cdot)$ are 1D-RGNN layers each requiring FLOPs of ${\cal O}(L_F N^2 J_{\ell}^2)$. Again, since $q_{\sf R}({\bf d}_k^{(\ell-1)}, {\bf d}_j^{(\ell-1)}), j=1,\cdots,K$ need to be computed to obtain ${\bf d}_k^{(\ell)}$, and ${\bf d}_k^{(\ell)}, k=1,\cdots,K$ need to be computed in the $\ell$-th layer of 2D-RGNN, the number of FLOPs for the $\ell$-th 2D-RGNN layer is ${\cal O}(L_F K^2 N^2 J_{\ell}^2)$.
	
	With similar analysis, we can obtain that the number of FLOPs for computing the $\ell$-th 3D-RGNN layer when learning over the graph in section \ref{subsec:3d-vgnn} is ${\cal O}(L_F K^2 N_{RF}^2 N^2 J_{\ell}^2)$, and the number of FLOPs for computing the $\ell$-th $n$D-RGNN layer is ${\cal O}\big(L_F (\prod_{i=1}^n K_n^2) J_{\ell}^2\big)$.

\vspace{-3mm}\section{Simulation Results}\label{sec:simulation-results}\vspace{-1mm}
In this section, we evaluate the learning performance and validate the learning efficiency of the proposed RGNNs in terms of training complexity and size generalizability via simulations.

In addition to learning the precoding policies from problems \textbf{P1} and \textbf{P2}, we also consider an EE-maximal baseband precoding problem. We only evaluate the size-generalizability of the GNNs, because RGNNs are designed for enabling size-generalizability while the generalizability of GNNs to other environmental parameters such as SNR or channel distribution has been evaluated in \cite{ModelGNN_GJ_2022, lee2020wireless}.

\vspace{-3mm}\subsection{Learning Baseband Precoding} \label{sec:simu-fdp}
The training and test samples are generated from uncorrelated Rayleigh fading channel, i.e., the elements in ${\bf H}$ follow complex Gaussian distribution ${\cal CN}(0,1)$. The size of each sample of ${\bf H}$ depends on the considered settings of $N$ and $K$. We generate 100,000 samples for every pair of $N$ and $K$ as the training set but the samples actually used for training may be with a much smaller number. We generate another 1,000 samples as the test set.

To show how satisfying the three characteristics in (D1)$\sim$(D3) affects the learning performance, we compare the \emph{2D-RGNN} proposed in section \ref{subsec:2d-rgnn} with the following GNNs, which are also used to learn ${\mathbf{W}}_{BB}^{\star}=F_b({\mathbf{H}})$.
\begin{itemize}
	\item \emph{2D-Vanilla-GNN}: This is the GNN designed in \cite{ZBC_WCNC}, which was reviewed in section \ref{sec:2d-vgnn}.
	\item \emph{Edge-GAT}: This is the extended version of the edge-GAT designed in \cite{Edge-GAT}, which was introduced in section \ref{sec:edge-gat}.
\item \emph{Model-GNN}: This is a model-driven GNN proposed in \cite{ModelGNN_GJ_2022} where the mathematical model of matrix pseudo-inverse was integrated into the GNN, which can also well-learn baseband precoding with various objectives and constraints.
\end{itemize}
The performance of GNN-N and GNN-P is similar to the performance of the 2D-Vanilla-GNN, hence is not provided.

The fine-tuned hyper-parameters of the GNNs are provided in Table \ref{table: hyper params}, where $J_{\ell}$ is the number of elements in the hidden representation vector in the $\ell$-th layer.
Since these parameters are similar for different settings of $K$ and $N$, we use the same hyper-parameters for all values of $K$ and $N$. The DNNs are trained with unsupervised deep learning \cite{sun2019pimrc} by Adam algorithm \cite{dlbook}.

\begin{table}[!htb]
	\centering
	\caption{Hyper-parameters for GNNs.}\label{table: hyper params}\vspace{-2mm}
	\small
	\begin{tabular}{c|c|c|c|c|c}
		\hline\hline
		\multicolumn{2}{c|}{Network} & \tabincell{c}{Num. of \\hidden layers} & \tabincell{c}{Value of\\ $J_{\ell}$ in layers} & \tabincell{c}{Learning\\rate} & \tabincell{c}{Activation function\\ of hidden layers}  \\ \hline
		\multicolumn{2}{c|}{2D-RGNN}							   &	5	& [16, 32, 32, 32, 16]& 0.001 & Tanh	%($y=\frac{\exp(x)-\exp(-x)}{\exp(x)+\exp(-x)}$)		
		\\ \hline
		\multicolumn{2}{c|}{2D-Vanilla-GNN}							 &	  4	  & [64, 512, 512, 64]	& 0.001	& Relu %($y=\max(x,0)$)	
		\\ \hline
		\multicolumn{2}{c|}{Model-GNN}							 &	  4	  & [32, 32, 32, 8]	& 0.001	& ${\bf X}/\|{\bf X}\|_F$		\\ \hline
		\multicolumn{2}{c|}{Edge-GAT}							 &	  4	  & [16, 16, 4, 2]	& 0.001	& ${\bf X}/\|{\bf X}\|_F$		\\ \hline
		\hline
	\end{tabular}
\end{table}

The performance of each DNN is obtained by averaging the tested results with five independently trained DNNs to reduce the impact of randomness caused by the selected samples  and initial weights. Each trained DNN is obtained with $N_{\sf tr}$ samples randomly selected from the training set and then tested on 1,000 samples.

\subsubsection{Learning to Maximize SE}
We first evaluate the performance of the GNNs for learning the precoding policy from problem \textbf{P1}. The activation function of the output layer of the GNNs is set as $\sigma({\bf X})=\frac{\bf X}{\|{\bf X}\|_F}\cdot P_{\max}$ to satisfy the power constraint in \eqref{eq:bb-constraint}. The learning performance is measured by the SE ratio, i.e., the ratio of the SE achieved by the learned policy to the SE achieved by a numerical algorithm, which is selected as the WMMSE algorithm.

\begin{figure}[!htb]
	\centering
	\begin{minipage}[t]{0.45\linewidth}	
		\subfigure[SNR = 10 dB]{
			\includegraphics[width=\textwidth]{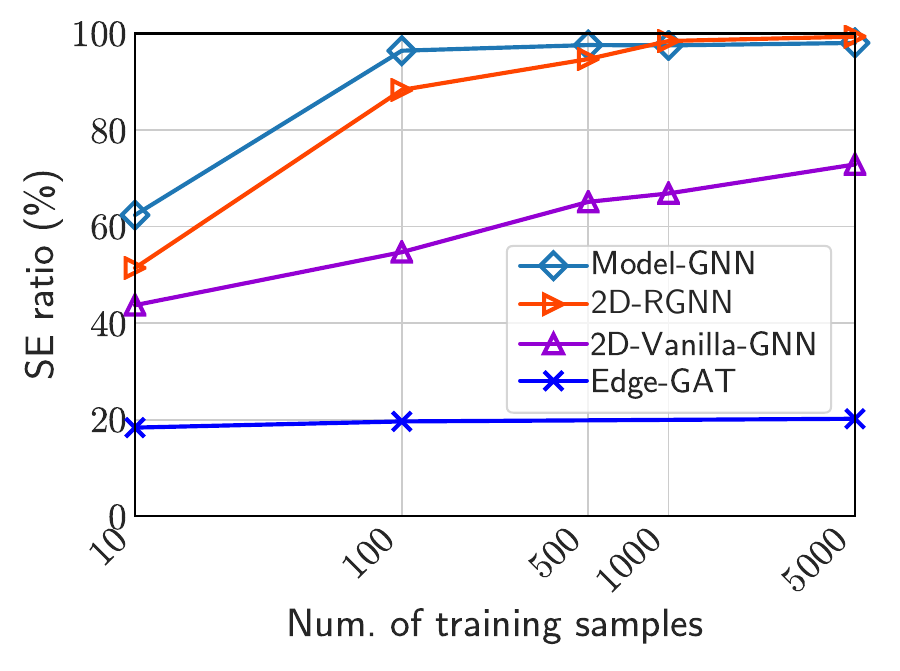}}
	\end{minipage}
	\begin{minipage}[t]{0.45\linewidth}	
		\subfigure[SNR = 20 dB]{
			\includegraphics[width=\textwidth]{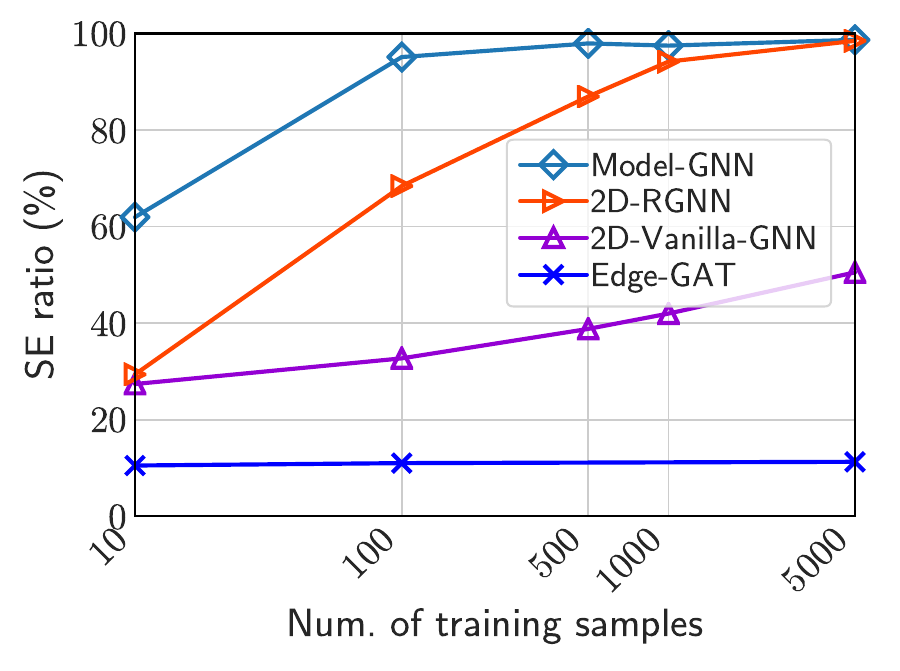}}
	\end{minipage}
	\vspace{-2mm}
	\caption{Learning performance, $N=8,K=4$. The SE ratio of the WMMSE algorithm is 100\%.}\label{fig-SE-Ntr-N8K4}\vspace{-1mm}
\end{figure}

In Fig. \ref{fig-SE-Ntr-N8K4}, we provide the SE ratios when the problem scales (in particular, the values of $K$ and $N$) are identical in training and test samples. It can be seen that the RGNN, which is purely data-driven, can achieve higher than $95\%$ ratio with 1,000 samples. The Model-GNN achieves a higher SE ratio than the RGNN with fewer training samples.
%, because the mapping that needs to be learned from data becomes simpler due to incorporating with mathematical model.
However, the Model-GNN cannot be used to learn other precoding policies such as hybrid precoding, because the introduced model is tailored to the baseband precoding. The SE ratios of all the GNNs become higher when SNR is low.
The 2D-Vanilla-GNN and edge-GAT do not perform well with less than 5,000 training samples. This is because the number of training samples required for them to achieve a satisfactory performance grows with $K$ according to Remark \ref{remark:gat-cplxty}, and 5,000 training samples are not sufficient for $K=4$.\footnote{According to the results in \cite{ZBC_WCNC}, 200,000 samples are required for the 2D-Vanilla-GNN to achieve a 99\% sum rate ratio when $N=8$ and $K=4$,  SNR$=10$ dB.}
We do not show the performance of the extended edge-GAT in the sequel because it does not perform well.

\begin{figure}[!htb]
	\centering
	\includegraphics[width=.7\linewidth]{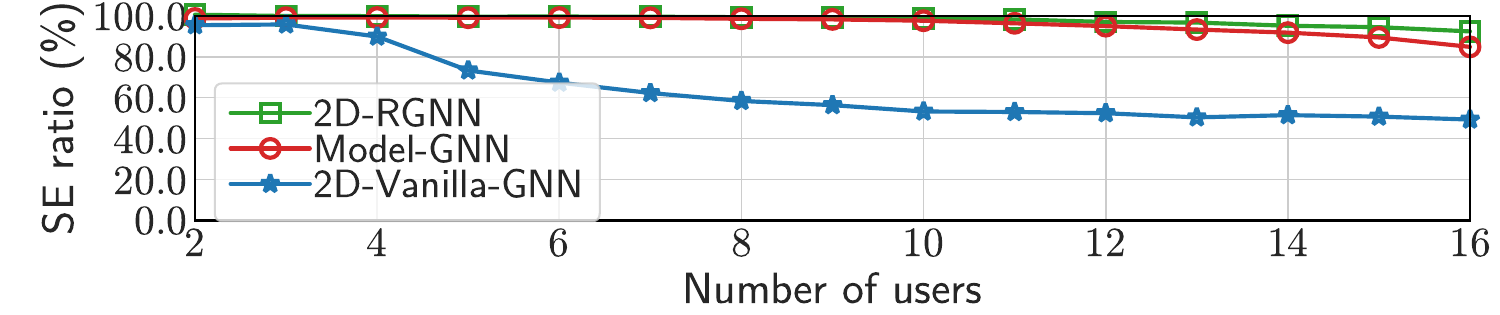}\vspace{-2mm}
	\caption{Generalization ability of GNNs to $K$, $N=16$, SNR = 10 dB.}
	\label{fig-scale-SE}
\end{figure}

In Fig. \ref{fig-scale-SE}, we show the generalizability of the GNNs to the number of users. The GNNs are trained with 1,000 samples, which are generated in the scenario where $K$ follows two-parameter exponential distribution with mean value of $5$ and standard deviation of $3$. In this case, $90$\% of samples are with $K<8$. The test samples are generated in the scenario where $K\thicksim{\mathbb U}(2,16)$, and ${\mathbb U}(\cdot,\cdot)$ stands for uniform distribution.
We can see that the SE ratios of the RGNN and Model-GNN  degrade very slowly with $K$, which validates their generalizability to $K$.

In Table \ref{table: train cplxty}, we provide the sample complexity, time complexity and space complexity for training the GNNs, which are respectively the minimal number of training samples, the time used for training, and the number of trainable parameters for achieving an expected performance (set as 95\% SE ratio).
All the simulation results are obtained on a computer with one 14-core Intel i9-9940X CPU, one Nvidia RTX 2080Ti GPU, and 64 GB memory.

\begin{table}[!htb]
	\renewcommand{\arraystretch}{1.2}
	\setlength\tabcolsep{3.0pt}
	\centering
	\small
	\caption{Training Complexity of GNNs, SNR=10 dB} \label{table: train cplxty}\vspace{-2mm}
	\begin{tabular}{c|c|c|c|c}
		\hline \hline
		&              & \bf 2D-RGNN & \bf Model-GNN & \bf 2D-Vanilla-GNN  \\ \hline
		\multirow{3}{*}{\tabincell{c}{$N=8$\\$K=4$}}   & Sample       & 500 & 100       & 40,000          \\ \cline{2-5}
		& Time  & 7 min 2 s & 32 s     & 43 min 16 s     \\ \cline{2-5}
		& Space  & 55,476 & 52,946 & 1,315,206   \\ \hline
		\multirow{3}{*}{\tabincell{c}{$N=16$\\$K=8$}}   & Sample       & 300 & 50       & $>$100,000*            \\ \cline{2-5}
		& Time  & 4 min 16 s & 18 s     & $>$3 hours          \\ \cline{2-5}
		& Space  & 55,476 & 52,946 & $>$1,315,206 \\ \hline
		\multirow{3}{*}{\tabincell{c}{$N=32$\\$K=16$}} & Sample     & 200  & 100       &   $>$100,000           \\ \cline{2-5}
		& Time & 3 min 52 s & 39 s     &  $>$6 hours    \\
		\cline{2-5}
		& Space  & 55,476 & 52,946 & $>$1,315,206   \\ \hline
		\hline
	\end{tabular}
	\flushleft{*: The ``$>$'' means that the 2D-Vanilla-GNN cannot achieve the expected performance when using the number of training samples, training time and the number of trainable parameters provided.}
\end{table}

As expected, the Model-GNN has the lowest training complexity  because the mapping that needs to be learned is simpler. Although the sample and time complexities of the RGNN are higher than the Model-GNN, they are affordable, because the RGNN is size-generalizable and does not need to be re-trained when the number of users in the system changes. The 2D-RGNN has a much lower training complexity than the 2D-Vanilla-GNN, especially when the problem scale is large. Moreover, the training complexity of the RGNN decreases with $K$, which validates the analysis in section \ref{sec:rgnn-complexity}.

\subsubsection{Learning to Maximize EE}
To demonstrate that the proposed RGNN can also learn other baseband precoding policies, we consider the following optimization problem,
\begin{subequations}\label{P2: EE max}
	\begin{align}
	\textbf{P3:}~~\max_{{\bf W}_{BB}}~~ & \frac{{\sum_{k=1}^K R_k({\bf H }, {\bf W}_{BB})}}{{\rho {\sf Tr}({\bf W}_{BB}^{\sf H}{\bf W}_{BB}) + NP_c + P_0}}, \label{P2: obj} \\
	{\rm s.t.}~~ & R_k({\bf H}, {\bf W}_{BB}) \geq R_{\min}, k=1,\cdots,K, \label{P2: constraint}
	\end{align}
\end{subequations}
where \eqref{P2: obj} is the EE, $R_k({\bf H }, {\bf W}_{BB})$ is the data rate of UE$_k$ defined in \eqref{eq:bb-object}, $P_c$  is the circuit power consumption of each antenna, $P_0$ is the constant power consumed at the BS, $1/\rho$ is the power amplifier efficiency, \eqref{P2: constraint} is the quality of service (QoS) constraint, and $R_{\min}$ is the minimal data rate requirement of each user. In the simulation, $P_c=17.6$ W, $P_0=43.3$ W, $1/\rho=0.311$, which come from \cite{auer2011much} for a macro BS.

The EE-maximal problem can be solved with the numerical algorithm proposed in \cite{LY_TVT_2015}, which is referred to as \emph{EE-numerical} in the sequel. We set the number of iterations as 100 such that its computation time is affordable.
The algorithm can only find local optimal solutions after iterations due to the non-convexity of the problem.

 Since the QoS constraints cannot be satisfied by a normalization layer as for problem \textbf{P1}, we train the GNNs with unsupervised deep learning using the Lagrangian multiplier method \cite{sun2019pimrc}, which strives to satisfy the QoS requirements during training but cannot ensure the constraints to be  satisfied. Hence, we measure the learning performance in terms of EE and QoS satisfaction by \emph{EE ratio} (EER) and \emph{Constraint Satisfaction Ratio} (CSR), which are averaged over all the test samples. The EER is the ratio of EE achieved by the learned policy to the EE achieved by \emph{EE-numerical}. The CSR is  the percentage of the users whose QoS requirements are satisfied.

\begin{figure}[!htb]
	\centering
	\includegraphics[width=.7\linewidth]{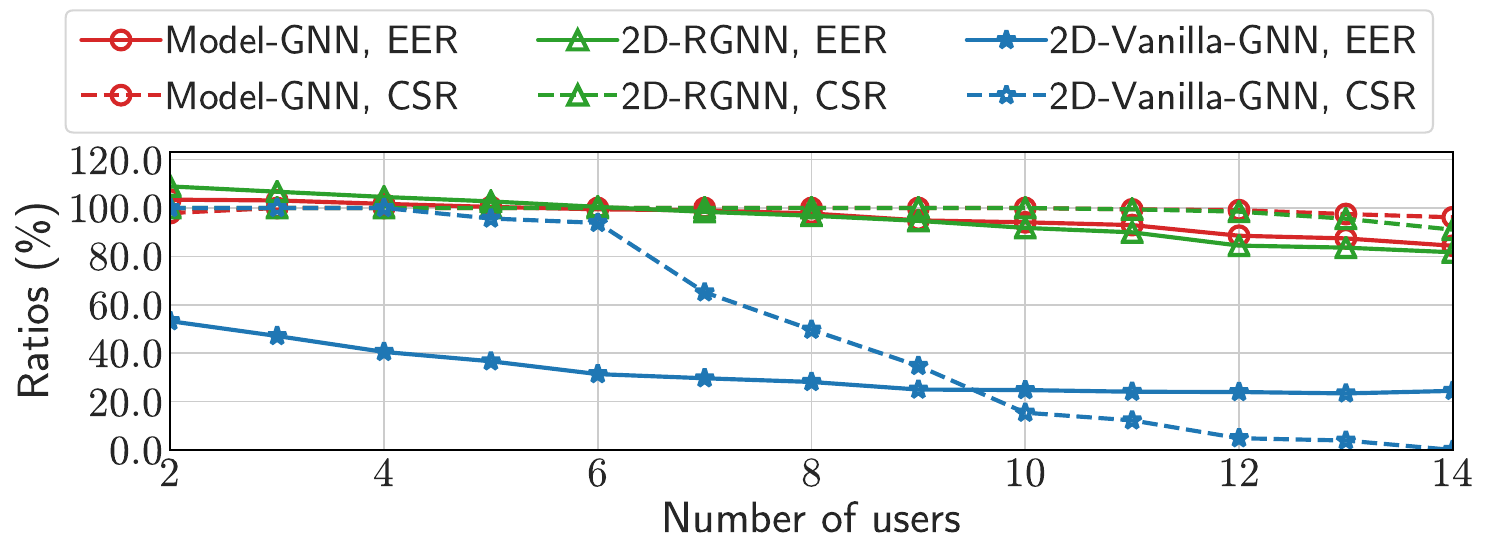}\vspace{-2mm}
	\caption{Generalization ability of the GNNs to $K$, $N=16$, $R_{\min}=2$ bps/Hz, SNR = 10 dB.}
	\label{fig-scale-EE}
\end{figure}

In Fig. \ref{fig-scale-EE}, we show the EER and CSR achieved by the GNNs tested in scenarios with different values of $K$ from the training samples. The training and test samples are the same as those generated for the results in Fig. \ref{fig-scale-SE}. It can be seen that both ratios of the RGNN and Model-GNN degrade slowly with $K$, which demonstrates their superior generalizability.

\vspace{2mm}

Our simulation results show that both the RGNN and Model-GNN can also be well-generalized to the number of antennas, which are not provided for conciseness.

\vspace{-3mm}\subsection{Learning Hybrid Precoding}\vspace{-1mm}
In this subsection, we take learning the hybrid precoding policy to maximize SE (i.e., from problem \textbf{P2}) as an example to evaluate the learning performance and size-generalizability of the proposed 3D-RGNN. In the output layer of the 3D-RGNN, after decreasing the dimension of ${\bf D}^{(L)}$ to two matrices ${\bf D}^{RF}$ and ${\bf D}^{BB}$ with the method in \cite{LSJ_MultiDim_GNN_2022}, the following activation functions are applied to satisfy the constraints \eqref{eq:constraint3} and \eqref{eq:constraint2}, i.e., $({\bf W}^{RF})_{j,l}=({\bf D}^{RF})_{j,l}/|({\bf D}^{RF})_{j,l}|$ and ${\bf W}^{BB}={\bf D}^{BB}/\|{\bf W}^{RF}{\bf D}^{BB}\|_F$.

Since hybrid precoding is often used in mmWave systems with large number of antennas to reduce hardware complexity,
the training and test samples are generated from the Saleh-Valenzuela channel model that is usually used for mmWave channels. We consider uniform linear antenna array. The channel from the BS to one user can be modeled as ${\mathbf{h}} = \sqrt{\frac{N_t}{N_{cl}N_{ray}}}\sum_{i=1}^{N_{cl}}\sum_{j=1}^{N_{ray}}\alpha_{i,j} {\mathbf{f}}(\theta_{i,j}) \in \mathbb{C}^{N}$ \cite{MO}, where $N_{cl}$ is the number of scattering clusters, $N_{ray}$ is the number of scattering rays, $\alpha_{i,j}$ is a complex gain, and ${\mathbf{f}}(\theta_{i,j})$ is the array response. In the simulations, we set $N_{cl}=4$ and $N_{ray}=5$, and $\alpha_{i,j}\thicksim{\cal CN}(0, 1)$ \cite{MO}. The training and test samples are generated in the same way as in section \ref{sec:simu-fdp}.
\begin{table}[!htb]
	\centering
	\caption{Hyper-parameters of GNNs}\label{table: hyper params1}\vspace{-2mm}
	\small
	\begin{tabular}{c|c|c|c|c|c}
		\hline\hline
		\multicolumn{2}{c|}{Network} & \tabincell{c}{Num. of \\hidden layers} & \tabincell{c}{Num. of\\ neurons in layers} & \tabincell{c}{Learning\\ rate} & \tabincell{c}{Activation function\\ of hidden layers}  \\ \hline
		\multicolumn{2}{c|}{3D-RGNN}							   &	3	& [16, 16, 4]& 0.001 & Tanh			\\ \hline
		\multicolumn{2}{c|}{3D-Vanilla-GNN}							 &	  4	  & [64, 64, 64, 64]	& 0.001	& Relu		\\ \hline
		\multicolumn{2}{c|}{2D-Vanilla-GNN}							 &	  4	  & [64, 64, 64, 64]	& 0.001	& Relu		\\ \hline
		\hline
	\end{tabular}
\end{table}

Problem \textbf{P2} can be solved via the manifold optimization (MO) algorithm \cite{MO}. The learning performance is again measured by the SE ratio, which is the ratio of the SE achieved by the policy learned with each DNN to the SE achieved by the MO algorithm.

We compare the 3D-RGNN proposed in section \ref{subsec:3d-rgnn} with the following GNNs, which are also used to learn $\{{\mathbf{W}}_{RF}^{\star},{\mathbf{W}}_{BB}^{\star}\}=F_h({\mathbf{H}})$.
\begin{itemize}
	\item \emph{3D-Vanilla-GNN}: This is the GNN proposed in \cite{LSJ_MultiDim_GNN_2022}, which was reviewed in section \ref{subsec:3d-vgnn}.
	\item \emph{2D-Vanilla-GNN}: This is the GNN reviewed in section \ref{sec:2d-vgnn}, which is used to learn over a graph with three types of vertices  (i.e., UEs, ANs and RFs) and edges between every two types of vertices. \end{itemize}
We also show the performance of the orthogonal matching pursuit (OMP) algorithm proposed in \cite{OMP}, which is also a numerical algorithm but with low computational complexity. We no longer compare with the Model-GNN since it cannot be used to learn hybrid precoding policy.

The fine-tuned hyper-parameters of the GNNs are shown in Table. \ref{table: hyper params1}.

\begin{figure}[!htb]
	\centering
	\begin{minipage}[t]{0.45\linewidth}	
		\subfigure[SNR = 10 dB]{
			\includegraphics[width=\textwidth]{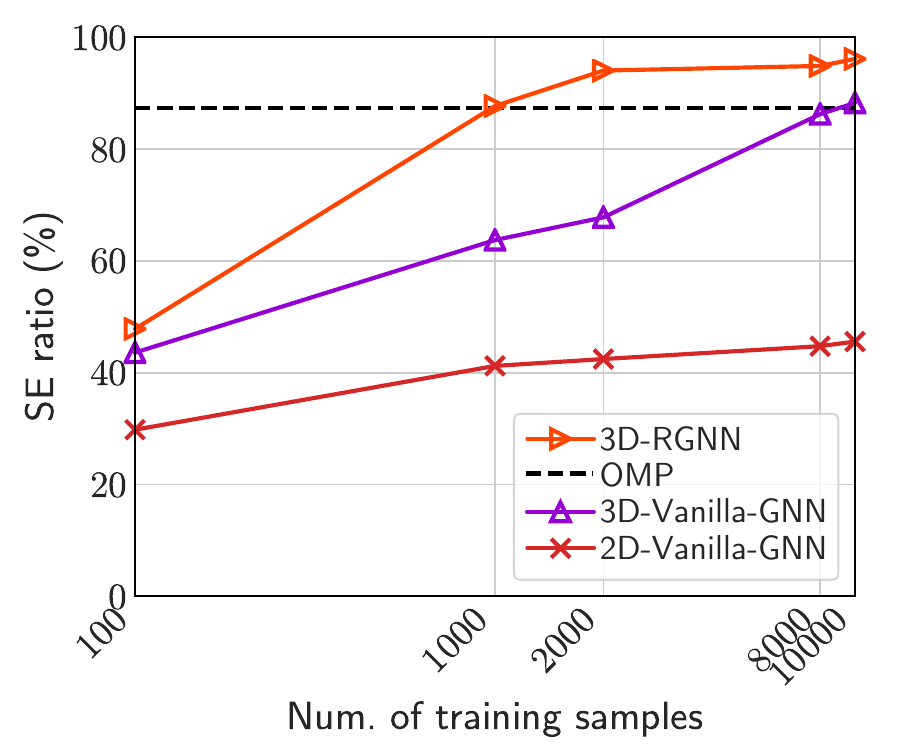}}
	\end{minipage}
	\begin{minipage}[t]{0.45\linewidth}	
		\subfigure[SNR = 20 dB]{
			\includegraphics[width=\textwidth]{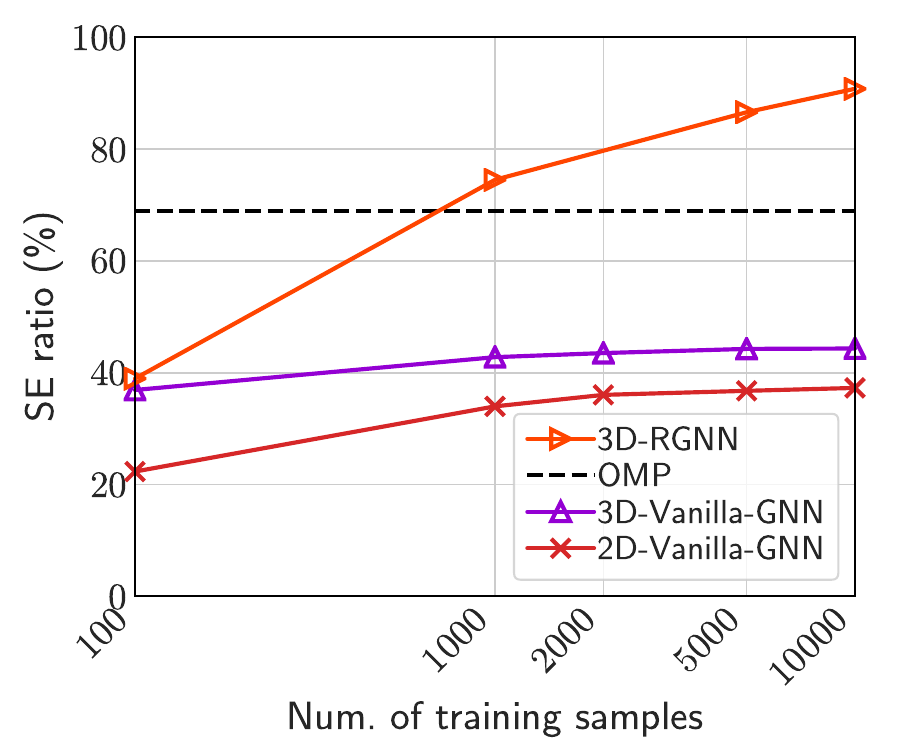}}
	\end{minipage}
	\vspace{-2mm}
	\caption{Learning performance, $N=8, N_s=6, K=3$. The SE ratio of the MO algorithm is 100\%.}\label{fig:hbp-perf}\vspace{-1mm}
\end{figure}

In Fig. \ref{fig:hbp-perf},
we show the SE ratios achieved by the GNNs and the OMP algorithm. When SNR is 20 dB, the GNNs are firstly trained under SNR = 10 dB and then fine-tuned under SNR = 20 dB to avoid the convergence to bad local minima.
%By unsupervisely training the GNNs with loss function of \eqref{eq:obj}, this can be accomplished by setting $P_{\max}/\sigma_0^2$ as $10$ and $100$ during pre-training and fine-tuning, respectively.
It is shown that the 3D-RGNN performs much better than the 3D-Vanilla-GNN, and can achieve near 90\% SE ratio with only 1,000 samples when SNR is 10 dB. The 3D-RGNN outperforms the OMP algorithm with more than 1,000 samples. The 2D-Vanilla-GNN does not perform well due to information loss, as mentioned in section \ref{subsec:3d-vgnn}. All the GNNs perform better when SNR is lower, because in this case the policy becomes smoother and is easier to learn \cite{ZBC_WCNC}.

\begin{figure}[!htb]
	\centering
	\includegraphics[width=.7\linewidth]{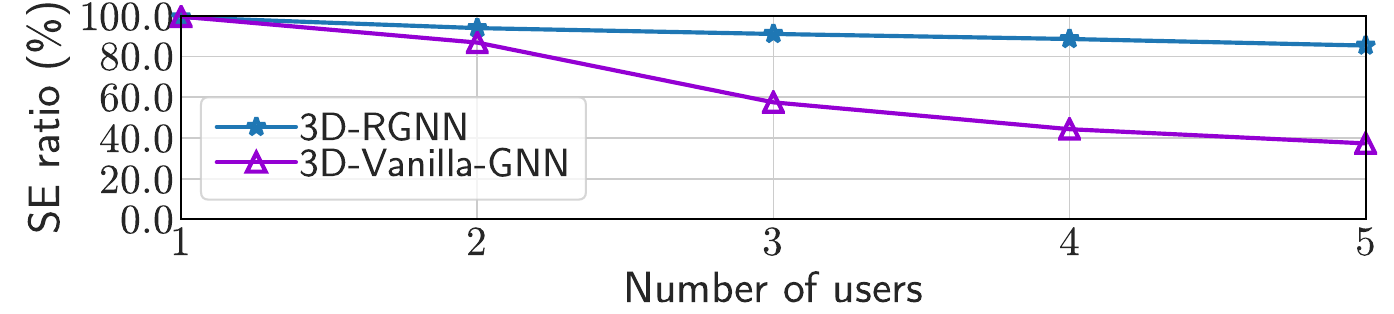}
	\caption{Generalization ability of the GNNs to $K$,  $N=16$ and $N_s=6$, SNR = 10 dB.}
	\label{fig-hbp-scale-SE}
\end{figure}

In Fig. \ref{fig-hbp-scale-SE}, we evaluate the size-generalization ability of the GNNs. The result of the 2D-Vanilla-GNN is not provided, since it cannot perform well even when the test samples are generated with the same distribution as the training samples as shown in Fig. \ref{fig:hbp-perf}. The GNNs are trained with 1,000 samples generated in scenarios where $K$ follows two-parameter exponential distribution with mean value of $3$ and standard deviation of $2$. The test samples are generated in scenarios where $K \thicksim{\mathbb U}(1,5)$. It can be seen that the SE ratios of the 3D-RGNN degrade much slower with $K$ than the 3D-Vanilla-GNN. This demonstrates the size-generalizability of the 3D-RGNN.

\vspace{2mm}

% The time required for training the 3D-RGNN and 3D-Vanilla-GNN in Fig. \ref{fig-scale-SE} is 15 h 9 min 6 s and 226 s, respectively.
	
	% The time required for training the 3D-RGNN and 3D-Vanilla-GNN in Fig. \ref{fig-scale-SE} is 15 h 9 min 6 s and 226 s, respectively.
	\section{Conclusions}\label{sec:conclusions}
	\vspace{-2mm}
	In this paper, we proposed an approach of designing size-generalizable GNNs for learning precoding policies. We first found that both the input-output relations of the iterations of numerical algorithms for optimizing baseband precoding and hybrid precoding can be expressed as a 1D-PE function, which is with a special structure and is one of the functions satisfying the 1D-PE property, and found that the re-expressed iterative equations of the algorithms can be regarded as the well-trained update equation of a GNN with size-generalizability (i.e., WGNN).
	Then, we compared the update equations of a 2D-Vanilla-GNN, an edge-GAT and the WGNN for learning a baseband precoding policy, which are with matched PE property to the policy. Based on the observation from the comparison, we explained why the 2D-Vanilla-GNN and the edge-GAT cannot be well-generalized to problem scales, and identified three key characteristics in addition to satisfying PE property for enabling GNNs to be size generalizable. We proceeded to design a RGNN with these characteristics and satisfies the high-dimensional PE property of a policy in a recursive manner. Interestingly, the sample complexity for training the RGNN decreases as the number of users
grows. The proposed RGNNs are applicable to learning precoding policies over hyper-graphs with multiple types of vertices, and the approach of designing the RGNNs can be used to design size-generalizable GNNs for other kinds of wireless policies. The superior performance of the RGNNs with respect to existing counterparts in terms of size generalizability and sample complexity was validated via simulations for learning baseband precoding and hybrid precoding policies.
	
	% The idea of designing size-generalizable GNNs from re-expressing the iterative equations of numerical algorithms is distinct from the idea of deep unfolding network, whose application is restricted by introducing the mathematical model of the numerical algorithms. The proposed idea can be applied to other problems, such as wide-band hybrid precoding in MIMO systems.

	\begin{appendices} \numberwithin{equation}{section}
		\renewcommand{\thesectiondis}[2]{\Alph{section}}

\section{Proof of Proposition \ref{prop:wmmse-iter-eq}}\label{appendix:precod-wmmse}
The precoding matrix can be optimized from the WMMSE algorithm iteratively as \cite{WMMSE2011Shi},
	\begin{eqnarray}
		{\bf w}_{BB_k}^{(\ell)} \!\!\!&\!\!\!=\!\!\!&\!\!\! \Big(\textstyle\sum_{j=1}^K z_j^{(\ell-1)} |u_j^{(\ell-1)}|^2 {\bf h}_j{\bf h}_j^{\sf H} + \mu\big(P_{\max}, \sum_{j=1}^K z_j^{(\ell-1)} |u_j^{(\ell-1)}|^2 {\bf h}_j{\bf h}_j^{\sf H}, \notag \\
		&& \hspace{10mm}\textstyle\sum_{j=1}^K (z_j^{(\ell-1)})^2 |u_j^{(\ell-1)}|^2 {\bf h}_j{\bf h}_j^{\sf H}\big)\cdot \mathbf{I}\Big)^{-1}\cdot{\bf h}_k u_k^{(\ell-1)} z_k^{(\ell-1)}, \label{eq:precod-v-upd}\\
		u_k^{(\ell)} \!\!&\!\!=\!\!&\!\! \frac{{\bf h}_k^{\sf H}{\bf w}_{BB_k}^{(\ell-1)}}{\sum_{j=1}^K |{\bf h}_k^{\sf H}{\bf w}_{BB_j}^{(\ell-1)}|^2 + \sigma_0^2}, \label{eq:precod-u-upd}\\
		z_k^{(\ell)} \!\!&\!\!=\!\!&\!\! \frac{1}{1-u_k^{(\ell-1)}{\bf h}_k^{\sf H}{\bf w}_{BB_k}^{(\ell-1)}}, \label{eq:precod-w-upd}
	\end{eqnarray}
	where ${\bf w}_{BB_k}^{(\ell)}$ is the updated precoding vector for the $k$-user in the $\ell$-th iteration, $u_k^{(\ell)}$ and $z_k^{(\ell)}$ are the intermediate variables for the $k$-user in the $\ell$-th iteration, $\mu(\cdot)$ is a function for computing Lagrange multiplier such that the power constraint \eqref{eq:bb-constraint} is satisfied, and $\bf I$ denotes unit matrix.

To prove the proposition, we first find all the variables for each user in each iteration that can be expressed as an updatable vector. Then, we find the summations in each iteration that can be regarded as information aggregation with $\sum_{j=1,j\neq k}^K q_{\sf BB}(\cdot)$. Finally, we re-express other operations except the summations as information combination with $f_{\sf BB}(\cdot)$.

\emph{(1) Finding the variables for the $k$-user:}
	By adding an auxiliary equation
	\begin{equation} \label{eq:precod-h-upd}
	{\bf h}_k ={\bf h}_k,
	\end{equation}
	and by denoting a vector for the $k$-user with $2N+2$ elements as,
	\begin{equation}\label{eq:upd-vec-bb-precod}
	{\bf d}_k^{(\ell)}= [{\bf d}_{k,1}^{(\ell)}, {\bf d}_{k,2}^{(\ell)}, {\bf d}_{k,3}^{(\ell)}, {\bf d}_{k,4}^{(\ell)}]\triangleq[{\bf w}_{BB_k}^{(\ell)}, u_k^{(\ell)}, z_k^{(\ell)}, {\bf h}_k],
	\end{equation}
	we can re-write \eqref{eq:precod-v-upd}$\sim$\eqref{eq:precod-h-upd} as an iterative equation of updating ${\bf d}_k^{(\ell)}$ with ${\bf d}_1^{(\ell-1)},\cdots, {\bf d}_K^{(\ell-1)}$.
% with ${\bf D}^{(\ell-1)}=[{\bf d}_1^{(\ell)},\cdots,{\bf d}_K^{(\ell)}]$.
	
	\emph{(2) Finding the aggregator:} There are three different summations in \eqref{eq:precod-v-upd}$\sim$\eqref{eq:precod-h-upd}. To show that they play the role of information aggregation, we express the terms in these summations as a multivariate non-linear non-element-wise function of ${\bf d}_k^{(\ell-1)}$ and ${\bf d}_j^{(\ell-1)}$,
	\begin{eqnarray}\label{eq:psi}
	&&\Big[ z_j^{(\ell-1)} |u_j^{(\ell-1)}|^2 {\bf h}_j{\bf h}_j^{\sf H},
	(z_j^{(\ell-1)})^2 |u_j^{(\ell-1)}|^2 {\bf h}_j{\bf h}_j^{\sf H},
	|{\bf h}_k^{\sf H}{\bf w}_{BB_j}^{(\ell-1)}|^2\Big]\notag\\
	\!\!&\!\!=\!\!&\!\!\Big[{\bf d}_{j,3}^{(\ell-1)} |{\bf d}_{j,2}^{(\ell-1)}|^2 {\bf d}_{j,4}^{(\ell-1)}{\bf d}_{j,4}^{(\ell-1){\sf H}},
	({\bf d}_{j,3}^{(\ell-1)})^2 |{\bf d}_{j,2}^{(\ell-1)}|^2 {\bf d}_{j,4}^{(\ell-1)}{\bf d}_{j,4}^{(\ell-1){\sf H}},
	| {\bf d}_{k,4}^{(\ell-1){\sf H}}{\bf d}_{j,1}^{(\ell-1)}|^2\Big]\notag\\
	\!\!&\!\!\triangleq\!\!&\!\! q_{\sf BB}({\bf d}_k^{(\ell-1)}, {\bf d}_j^{(\ell-1)}),
	\end{eqnarray}
where ${\bf d}_{k,i}^{(\ell-1)}$ is the $i$-th component of ${\bf d}_{k}^{(\ell-1)}$, for example, ${\bf d}_{k,1}^{(\ell-1)}={\bf w}_{BB_k}^{(\ell-1)}$.
%	These two terms corresponds to the information extraction and aggregation with processor and pooling function in the GNN.
%	
%	There are two summations respectively in \eqref{eq:precod-v-upd} and \eqref{eq:precod-u-upd}. Since the terms in the summations consist of the variables in ${\bf d}_j^{(\ell-1)}$ and  ${\bf d}_k^{(\ell-1)}$, the vector composed by the two terms is a function of ${\bf d}_j^{(\ell-1)}$ and  ${\bf d}_k^{(\ell-1)}$, i.e.,
%	\begin{eqnarray}\label{eq:psi}
%	\Big[ z_j^{(\ell-1)} |u_j^{(\ell-1)}|^2 {\bf h}_j{\bf h}_j^{\sf H},  |{\bf h}_k^{\sf H}{\bf w}_{BB_j}^{(\ell-1)}|^2\Big]\triangleq q_{\sf BB}({\bf d}_k^{(\ell-1)}, {\bf d}_j^{(\ell-1)}),
%	\end{eqnarray}
%	which is a multivariate non-linear non-element-wise function of ${\bf d}_k^{(\ell-1)}$ and ${\bf d}_j^{(\ell-1)}$. To show that the outputs are the functions of the input variables more clearly, we re-express \eqref{eq:psi} as,
%	\begin{equation}\label{eq:psi-1}
%	\Big[{\bf d}_{j,3}^{(\ell-1)} |{\bf d}_{j,2}^{(\ell-1)}|^2 {\bf d}_{j,4}^{(\ell-1)}{\bf d}_{j,4}^{(\ell-1){\sf H}},  | {\bf d}_{k,4}^{(\ell-1){\sf H}}{\bf d}_{j,1}^{(\ell-1)}|^2\Big]\triangleq q_{\sf BB}({\bf d}_k^{(\ell-1)}, {\bf d}_j^{(\ell-1)}),
%	\end{equation}
%	where ${\bf d}_{k,i}^{(\ell-1)}$ is the $i$-th component of ${\bf d}_{k}^{(\ell-1)}$, for example, ${\bf d}_{k,1}^{(\ell-1)}={\bf w}_{BB_k}^{(\ell-1)}$.
	The input and output sizes (i.e., the numbers of elements in the input and the output) of $q_{\sf BB}(\cdot)$ are respectively $4N+4$ and $N^2+2$. Then, the three summations in \eqref{eq:precod-v-upd} and \eqref{eq:precod-u-upd} can be written as
	\begin{equation}\label{eq:appendix-aggr-vec}
	{\bf e}_k^{(\ell)} = [{\bf e}_{k,1}^{(\ell)}, {\bf e}_{k,2}^{(\ell)}, {\bf e}_{k,3}^{(\ell)}] \triangleq \textstyle\sum_{j=1}^K q_{\sf BB}({\bf d}_k^{(\ell-1)},{\bf d}_j^{(\ell-1)}).
	\end{equation}
	
	\emph{(3) Finding the combiner: }
%After substituting \eqref{eq:appendix-aggr-vec} into \eqref{eq:precod-v-upd}$\sim$\eqref{eq:precod-h-upd}, the variables in the iterative equation are within ${\bf d}_k^{(\ell-1)}$ and ${\bf e}_k^{(\ell-1)}$. Hence,
With the expressions of ${\bf d}_k^{(\ell-1)}$, ${\bf e}_k^{(\ell-1)}$ and their components ${\bf d}_{k,i}^{(\ell-1)}$ and ${\bf e}_{k,i}^{(\ell-1)}$, we can re-write \eqref{eq:precod-v-upd}$\sim$\eqref{eq:precod-h-upd} as an iterative equation in compact form as follows,
	\begin{eqnarray}\label{eq:appendix-fw}
	{\bf d}_k^{(\ell)} \!\!&\!\!=\!\!&\!\! [{\bf w}_{BB_k}^{(\ell)}, u_k^{(\ell)}, z_k^{(\ell)}, {\bf h}_k] \notag\\
	\!\!&\!\!=\!\!&\!\!
	\Bigg[
	\Big({\bf e}_{k,1}^{(\ell-1)}+ \mu\big(P_{\max}, {\bf e}_{k,1}^{(\ell-1)}, {\bf e}_{k,2}^{(\ell-1)}\big)\Big)^{-1}{\bf d}_{k,4}^{(\ell-1)}{\bf d}_{k,2}^{(\ell-1)}{\bf d}_{k,3}^{(\ell-1)},\notag\\
	&&\frac{({\bf d}_{k,4}^{(\ell-1)})^{\sf H}{\bf d}_{k,1}^{(\ell-1)}}{{\bf e}_{k,3}^{(\ell-1)}+\sigma_0^2},
	\frac{1}{1-{\bf d}_{k,2}^{(\ell-1)}({\bf d}_{k,4}^{(\ell-1)})^{\sf H}{\bf d}_{k,1}^{(\ell-1)}},
	{\bf d}_{k,4}^{(\ell-1)}
	\Bigg]\notag\\
	\!\!&\!\!\triangleq\!\!&\!\! f_{\sf BB}'({\bf d}_k^{(\ell-1)}, {\bf e}_k^{(\ell-1)})=\textstyle f_{\sf BB}'\Big({\bf d}_k^{(\ell-1)}, \sum_{j=1}^K q_{\sf BB}({\bf d}_k^{(\ell-1)},{\bf d}_j^{(\ell-1)})\Big)
	\notag\\
	\!\!&\!\!=\!\!&\!\! \textstyle f_{\sf BB}'\Big({\bf d}_k^{(\ell-1)}, q_{\sf BB}({\bf d}_k^{(\ell-1)},{\bf d}_k^{(\ell-1)}) + \sum_{j=1,j\neq k}^K q_{\sf BB}({\bf d}_k^{(\ell-1)},{\bf d}_j^{(\ell-1)})\Big) \notag\\
	\!\!&\!\!\triangleq\!\!&\!\! \textstyle f_{\sf BB}\Big({\bf d}_k^{(\ell-1)}, \sum_{j=1,j\neq k}^K q_{\sf BB}({\bf d}_k^{(\ell-1)},{\bf d}_j^{(\ell-1)})\Big).
	\end{eqnarray}
%	where $(a)$ comes from putting $q_{\sf BB}({\bf d}_k^{(\ell)},{\bf d}_k^{(\ell)})$ outside the summation term.

We can see from \eqref{eq:appendix-fw} that $f_{\sf BB}(\cdot)$ is a non-linear non-element-wise function of ${\bf d}_k^{(\ell-1)}$ and $\sum_{j=1,j\neq k}^K q_{\sf BB}({\bf d}_k^{(\ell-1)},{\bf d}_j^{(\ell-1)})$ with two input variables and four output variables.
	The input size of $f_{\sf BB}(\cdot)$ is $N^2+2N+4$, and the output size is $2N+2$.

\section{Proof of Appendix \ref{prop:wmmse-mo-iter-eq}}\label{appendix:hybrid-precod}
We prove that the iterative equation for optimizing hybrid precoding can be re-expressed as \eqref{eq:iter-hybrid-precod}, by following the three steps as in Appendix \ref{appendix:precod-wmmse}.

The hybrid precoding can be optimized from the WMMSE-MO algorithm iteratively as \cite{Model_DL_HybPrec_TCOM2023},
\begin{eqnarray}
	{\bf w}_{BB_k}^{(\ell)} \!\!&\!\!=\!\!&\!\! \Bigg(\sum_{j=1}^K z_j^{(\ell-1)}|u_j^{(\ell-1)}|^2 {\bf W}_{RF}^{(\ell-1)\sf H}{\bf h}_j{\bf h}_j^{\sf H}{\bf W}_{RF}^{(\ell-1)} +  \notag\\
	&& \frac{\sigma_0^2}{P_{\max}}\sum_{j=1}^K z_j^{(\ell-1)}|u_j^{(\ell-1)}|^2 {\bf W}_{RF}^{(\ell-1)\sf H}{\bf W}_{RF}^{(\ell-1)}\Bigg)^{-1}\!\!\cdot{\bf W}_{RF}^{(\ell-1)\sf H}{\bf h}_k u_k^{(\ell-1)} z_k^{(\ell-1)}\!, \label{eq:hybp-wbb-upd}\\
	u_k^{(\ell)} \!\!&\!\!=\!\!&\!\! \frac{{\bf h}_k^{\sf H}{\bf W}_{RF}^{(\ell-1)}{\bf w}_{BB_k}^{(\ell-1)}}{\sum_{j=1}^K |{\bf h}_k^{\sf H}{\bf W}_{RF}^{(\ell-1)}{\bf w}_{BB_j}^{(\ell-1)}|^2 + \frac{\sigma_0^2}{P_{\max}}\sum_{j=1}^K \|{\bf W}_{RF}^{(\ell-1)}{\bf w}_{BB_j}^{(\ell-1)}\|^2},\notag\\ \label{eq:hybp-u-upd}\\
	z_k^{(\ell)} \!\!&\!\!=\!\!&\!\! \frac{1}{1-u_k^{(\ell-1)}{\bf h}_k^{\sf H}{\bf W}_{RF}^{(\ell-1)}{\bf w}_{BB_k}^{(\ell-1)}}, \label{eq:hybp-z-upd}\\
	{\bf W}_{RF}^{(\ell)} \!\!&\!\!=\!\!&\!\! {\cal P}\Big({\bf W}_{RF}^{(\ell-1)} - \alpha\cdot{\rm grad} f({\bf W}_{RF}^{(\ell-1)})\Big)\label{eq:hybp-wrf-upd}\\
	{\rm grad} f({\bf W}_{RF}^{(\ell-1)})\!\!&\!\!=\!\!&\!\! \nabla f({\bf W}_{RF}^{(\ell-1)}) - {\cal R}\{\nabla f({\bf W}_{RF}^{(\ell-1)})\odot {\bf W}_{RF}^{(\ell-1)*}\}\odot{\bf W}_{RF}^{(\ell-1)}, \label{eq:hybp-grad-upd}\\
	\nabla f({\bf W}_{RF}^{(\ell-1)}) \!\!&\!\!=\!\!&\!\! \sum_{j=1}^K u_j^{(\ell-1)} z_j^{(\ell-1)}{\bf h}_j  {\bf W}_{RF}^{(\ell-1)} {\bf w}_{BB_j}^{(\ell-1)\sf H}
	+ \notag\\
	&&\sum_{j=1}^K |u_j^{(\ell-1)}|^2 z_j^{(\ell-1)} {\bf h}_j {\bf h}_j^{\sf H} {\bf W}_{RF}^{(\ell-1)} \!\!\sum_{m=1}^K {\bf w}_{BB_m}^{(\ell-1)}({\bf w}_{BB_m}^{(\ell-1)})^{\sf H}
	\!+ \notag\\
	&&\frac{\sigma_0^2}{P_{\max}} \!\!\sum_{j=1}^K |u_j^{(\ell-1)}|^2 z_j^{(\ell-1)} {\bf W}_{RF}^{(\ell-1)} \!\!\sum_{m=1}^K {\bf w}_{BB_m}^{(\ell-1)}({\bf w}_{BB_m}^{(\ell-1)})^{\sf H},
	\label{eq:hybp-nabla-upd}
\end{eqnarray}
where ${\bf w}_{BB_k}^{(\ell)}$, $u_k^{(\ell)}$ and $z_k^{(\ell)}$ are respectively the updated baseband precoding vector and the intermediate variables for UE$_k$ in the $\ell$-th iteration, ${\bf W}_{RF}^{(\ell)}$ is the updated analog precoding vector in the $\ell$-th iteration, ${\cal P}(\cdot)$ denotes the projection operation onto the unit modulus constraint, $\alpha$ is the step size for gradient descent, ${\rm grad}(\cdot)$ denotes the Riemannian gradient, $f(\cdot)$ denotes the objective function of problem \textbf{P2}, ${\cal R}(\cdot)$ denotes the operation of taking the real part of a complex value, and $\odot$ denotes Hadamard product.

\emph{(1) Finding the variables for the $k$-user:} By adding an auxiliary equation
\begin{equation}\label{eq:hybp-h-upd}
	{\bf h}_k = {\bf h}_k,
\end{equation}
and by denoting ${\bf d}_{nmk}^{(\ell)}\triangleq[w_{BB_{mk}}^{(\ell)}, u_k^{(\ell)}, z_k^{(\ell)}, {\bf W}_{RF_{mn}}^{(\ell)}, h_{nk}]$, we can express \eqref{eq:hybp-wbb-upd}$\sim$\eqref{eq:hybp-h-upd} as a iterative equation of updating ${\bf D}_k^{(\ell)}$ with ${\bf D}^{(\ell-1)}=[{\bf D}_1^{(\ell)},\cdots,{\bf D}_K^{(\ell)}]$, where
\begin{equation}\label{eq:upd-mat-hyb-precod}
	{\bf D}_k^{(\ell)}=[{\bf D}_{k,1}^{(\ell)},{\bf D}_{k,2}^{(\ell)},{\bf D}_{k,3}^{(\ell)},{\bf D}_{k,4}^{(\ell)},{\bf D}_{k,5}^{(\ell)}]\triangleq [{\bf w}_{BB_k}^{(\ell)}, u_k^{(\ell)}, z_k^{(\ell)}, {\bf W}_{RF}^{(\ell)}, {\bf h}_k],
\end{equation}
is a matrix with the element in the $n$-th row and the $m$-th column being ${\bf d}_{nmk}^{(\ell)}$.

\emph{(2) Finding the aggregator:} There are eight different summations in \eqref{eq:hybp-wbb-upd}$\sim$\eqref{eq:hybp-h-upd}. Since the terms in these summations are the variables in ${\bf D}_j^{(\ell-1)}$ and  ${\bf D}_k^{(\ell-1)}$, the vector composed by the eight terms is a function of ${\bf D}_k^{(\ell-1)}$ and ${\bf D}_j^{(\ell-1)}$, i.e.,
\begin{eqnarray}\label{eq:q_HB}
	&&\Bigg[
	z_j^{(\ell-1)}|u_j^{(\ell-1)}|^2 {\bf W}_{RF}^{(\ell-1)\sf H}{\bf h}_j{\bf h}_j^{\sf H}{\bf W}_{RF}^{(\ell-1)}, z_j^{(\ell-1)}|u_j^{(\ell-1)}|^2 {\bf W}_{RF}^{(\ell-1)\sf H}{\bf W}_{RF}^{(\ell-1)},  \notag\\
	&&\hspace{3mm}|{\bf h}_k^{\sf H}{\bf W}_{RF}^{(\ell-1)}{\bf w}_{BB_j}^{(\ell-1)}|^2, \|{\bf W}_{RF}^{(\ell-1)}{\bf w}_{BB_j}^{(\ell-1)}\|^2,  u_j^{(\ell-1)} z_j^{(\ell-1)}{\bf h}_j  {\bf W}_{RF}^{(\ell-1)} {\bf w}_{BB_j}^{(\ell-1)\sf H},   \notag\\
	&&\hspace{3mm}|u_j^{(\ell-1)}|^2 z_j^{(\ell-1)} {\bf h}_j {\bf h}_j^{\sf H} {\bf W}_{RF}^{(\ell-1)}, {\bf w}_{BB_j}^{(\ell-1)}({\bf w}_{BB_j}^{(\ell-1)})^{\sf H},  |u_j^{(\ell-1)}|^2 z_j^{(\ell-1)} {\bf W}_{RF}^{(\ell-1)}\Bigg]\triangleq q_{\sf HB}({\bf D}_k^{(\ell-1)}, {\bf D}_j^{(\ell-1)}), \notag\\
\end{eqnarray}
which is a multivariate non-linear non-element-wise function  with two input variables and eight output variables. The input and output sizes of $q_{\sf HB}(\cdot)$ are respectively  $10NN_s$ and $3NN_s + N_s^2 + 2$.
%We can also re-express \eqref{eq:q_HB} to show that the outputs are the functions of the input variables more clearly, but the re-expressed function is not provided here for conciseness.
Then, the eight summations in \eqref{eq:hybp-wbb-upd}$\sim$\eqref{eq:hybp-h-upd} can be written as
\begin{equation}\label{eq:appendix-aggr-mat}
{\bf E}_k^{(\ell)}=[{\bf E}_{k,1}^{(\ell)},\cdots,{\bf E}_{k,8}^{(\ell)}] \triangleq \textstyle\sum_{j=1}^K q_{\sf HB}({\bf D}_k^{(\ell)},{\bf D}_j^{(\ell)}).
\end{equation}

\emph{(3) Finding the combiner: }  With the expressions of ${\bf D}_k^{(\ell-1)}$ and ${\bf E}_k^{(\ell-1)}$ and their components ${\bf D}_{k,i}^{(\ell-1)}$ and ${\bf E}_{k,i}^{(\ell-1)}$,
%After substituting \eqref{eq:appendix-aggr-mat} into \eqref{eq:hybp-wbb-upd}$\sim$\eqref{eq:hybp-h-upd}, the variables in the iterative equation are within ${\bf D}_k^{(\ell-1)}$ and ${\bf E}_k^{(\ell-1)}$.
by further substituting \eqref{eq:hybp-nabla-upd} into \eqref{eq:hybp-grad-upd} and substituting \eqref{eq:hybp-grad-upd} into \eqref{eq:hybp-wrf-upd},
we can re-write \eqref{eq:hybp-wbb-upd}$\sim$\eqref{eq:hybp-wrf-upd} and \eqref{eq:hybp-h-upd} as an iterative equation in compact form as follows,
% a non-linear non-element-wise function of ${\bf D}_k^{(\ell-1)}$ and ${\bf E}_k^{(\ell)}$ as,
\begin{eqnarray}\label{eq:f_HB}
{\bf D}_k^{(\ell)} \!\!&\!\!=\!\!&\!\! [{\bf w}_{BB_k}^{(\ell)}, {\bf W}_{RF}^{(\ell)}, u_k^{(\ell)}, z_k^{(\ell)}, {\bf h}_k]\notag\\
\!\!&\!\!=\!\!&\!\!
\Bigg[
\Big({\bf E}_{k,1}^{(\ell-1)}+\frac{\sigma_0^2}{P_{\max}}{\bf E}_{k,2}^{(\ell-1)}\Big)^{-1}{\bf D}_{k,4}^{(\ell-1)\sf H}{\bf D}_{k,5}^{(\ell-1)}{\bf D}_{k,2}^{(\ell-1)}{\bf D}_{k,3}^{(\ell-1)},\notag\\
&& \frac{({\bf D}_{k,5}^{(\ell-1)})^{\sf H}{\bf D}_{k,4}^{(\ell-1)}{\bf D}_{k,1}^{(\ell-1)}}{{\bf E}_{k,3}^{(\ell-1)}+\frac{\sigma_0^2}{P_{\max}}{\bf E}_{k,4}^{(\ell-1)}},
\frac{1}{1-{\bf D}_{k,2}^{(\ell-1)}{\bf D}_{k,5}^{(\ell-1)\sf H}{\bf D}_{k,4}^{(\ell-1)}{\bf D}_{k,1}^{(\ell-1)}}, \notag\\
&& {\cal P}\Big({\bf D}_{k,4}^{(\ell-1)}-\alpha \big({\bf B}_k^{(\ell-1)}-{\cal R}\{{\bf B}_k^{(\ell-1)}\odot {\bf D}_{k,4}^{(\ell-1)*}\}\odot {\bf D}_{k,4}^{(\ell-1)}\big)\Big), {\bf D}_{k,5}^{(\ell-1)}
\Bigg]\notag\\
\!\!&\!\!\triangleq\!\!&\!\! f_{\sf HB}'({\bf D}_k^{(\ell-1)}, {\bf E}_k^{(\ell-1)})=\textstyle f_{\sf HB}'\Big({\bf D}_k^{(\ell-1)}, \sum_{j=1}^K q_{\sf HB}({\bf D}_k^{(\ell-1)},{\bf D}_j^{(\ell-1)})\Big)\notag\\
\!\!&\!\!=\!\!&\!\! \textstyle f_{\sf HB}'\Big({\bf D}_k^{(\ell-1)}, q_{\sf HB}({\bf D}_k^{(\ell-1)},{\bf D}_k^{(\ell-1)}) + \sum_{j=1,j\neq k}^K q_{\sf HB}({\bf D}_k^{(\ell-1)},{\bf D}_j^{(\ell-1)})\Big)\notag\\
\!\!&\!\!\triangleq\!\!&\!\! \textstyle f_{\sf HB}\Big({\bf D}_k^{(\ell-1)}, \sum_{j=1,j\neq k}^K q_{\sf HB}({\bf D}_k^{(\ell-1)},{\bf D}_j^{(\ell-1)})\Big),
\end{eqnarray}
where ${\bf B}_k^{(\ell-1)}\triangleq {\bf E}_{k,5}^{(\ell-1)} + {\bf E}_{k,6}^{(\ell-1)}{\bf E}_{k,7}^{(\ell-1)}+\frac{\sigma_0^2}{P_{\max}}{\bf E}_{k,8}^{(\ell-1)}{\bf E}_{k,7}^{(\ell-1)}$.
%, $(a)$ comes from putting $q_{\sf BB}({\bf D}_k^{(\ell)},{\bf D}_k^{(\ell)})$ outside the summation term.

We can see from \eqref{eq:f_HB} that $f_{\sf HB}(\cdot)$ is a non-linear non-element-wise function of ${\bf D}_k^{(\ell-1)}$ and $\sum_{j=1,j\neq k}^K q_{\sf HB}({\bf D}_k^{(\ell-1)},{\bf D}_j^{(\ell-1)})$. The input and output sizes of $f_{\sf HB}(\cdot)$ are respectively $13NN_s + N_s^2 + 2$ and $5NN_s$.

	\end{appendices}
	\bibliography{IEEEabrv,GJ}

\end{document}